\documentclass[epjc,aps,showpacs,preprintnumbers,amsmath,nofootinbib,amssymb]{revtex4}
\usepackage[dvips]{epsfig,subeqnarray}

\newcommand{\nc}[1]{\newcommand{#1}}

\nc{\its}[1]{\itshape #1 \upshape}
\nc{\mc}[3]{\multicolumn{#1}{#2}{#3}}

\nc{\bc}{\begin{center}}
\nc{\ec}{\end{center}}
\nc{\ig}[1]{\bc \includegraphics{#1} \ec}

\nc{\bo}[1]{\mbox{\boldmath \( #1 \! \! \)  \unboldmath}}
\nc{\beq}{\begin{equation}}
\nc{\eeq}{\end{equation}}
\nc{\bew}{\begin{eqnarray}}
\nc{\eew}{\end{eqnarray}}
\nc{\bs}{\begin{subeqnarray}}   
\nc{\es}{\end{subeqnarray}}     
\nc{\nnn}{\nonumber \\}
\nc{\Mvariable[1]}{\; #1}
\nc{\f}[2]{\frac{#1}{#2}}
\nc{\td}[2]{\f{d #1}{d #2}}
\nc{\pd}[2]{\f{\partial #1}{\partial #2}}
\nc{\suli}{\sum\limits}
\nc{\proli}{\prod\limits}
\nc{\ili}{\int\limits}
\nc{\sr}[2]{\stackrel{#1}{#2}}
\nc{\dps}{\displaystyle}
\nc{\ket}[1]{\left| #1 \right>}
\nc{\bra}[1]{\left< #1 \right|}
\nc{\bracket}[2]{\left< #1 \right| \left. \! #2 \right>}
\nc{\norm}[1]{\left\| #1 \right\|}
\nc{\lndm}[1]{\pd{^{#1} \ln{\det{M}}}{\mu^{#1}}}
\nc{\pdmm}[1]{M^{-1} \pd{^{#1} M}{\mu^{#1}}}
\nc{\pdm}{M^{-1}\pd{M}{\mu}}
\nc{\trac}[1]{\mbox{Tr}\left(#1\right)}

\nc{\muh}{\hat \mu}
\nc{\nuh}{\hat \nu}
\nc{\rhoh}{\hat \rho}
\nc{\sigmah}{\hat \sigma}

\def\lsim{\raise0.3ex\hbox{$<$\kern-0.75em\raise-1.1ex\hbox{$\sim$}}}
\def\gsim{\raise0.3ex\hbox{$>$\kern-0.75em\raise-1.1ex\hbox{$\sim$}}}

\begin{document}
\mbox{} \hfill BI-TP 2008/1\\[0mm]
\mbox{} \hfill BNL-NT-08/1\\[0mm]
\title{Lattice cut-off effects and their reduction 
in studies of QCD thermodynamics at non-zero temperature
and chemical potential 
       }

\author{P. Hegde,$^{\rm a,b}$, F. Karsch$^{\rm b,c}$, E. Laermann$^{\rm c}$ 
and S. Shcheredin$^{\rm c}$
       }

\address{
$^{\rm a}$Department of Physics and Astronomy, Stony Brook
University, Stony Brook, NY 11790, USA\\
$^{\rm b}$Physics Department, Brookhaven National Laboratory, 
Upton, NY 11973, USA \\
$^{\rm c}$Fakult\"at f\"ur Physik, Universit\"at Bielefeld, D-33615 Bielefeld,
Germany
}

\date{\today}

\begin{abstract}
We clarify the relation between the improvement of dispersion relations
in the fermion sector of lattice regularized QCD and the improvement of 
bulk thermodynamic 
observables. We show that in the infinite temperature limit the cut-off 
dependence in dispersion relations can be eliminated up to ${\cal O}(a^n)$
corrections, if the quark propagator is chosen to be rotationally invariant
up to this order. In bulk thermodynamic observables this eliminates cut-off 
effects up to the same order at vanishing as well as non-vanishing chemical
potential. We furthermore show, that in the infinite temperature, ideal gas 
limit the dependence of finite cut-off corrections on the chemical potential
is given by Bernoulli polynomials which are universal as they do not depend
on a particular discretization scheme. We explicitly calculate leading and
next-to-leading order cut-off corrections for some staggered and Wilson
fermion type actions and compare these with exact evaluations of the free
fermion partition functions. This also includes the chirally invariant 
overlap and domain wall fermion formulations.

\end{abstract}

\pacs{11.15.Ha, 11.10.Wx, 12.38Gc, 12.38.Mh}

\maketitle

\section{Introduction}
\label{intro}

Numerical studies of lattice regularized QCD face the
problem of lattice discretization errors which complicate the
extraction of physical results in the continuum limit. This 
problem has been addressed ever since the formulation of a systematic
improvement scheme for gauge theories by Symanzik \cite{Symanzik}.
Aside from the now widely used Symanzik improved gauge actions
it also led to tree-level improved actions for the fermion sector of 
QCD like the Naik \cite{Naik} and p4 \cite{p4action} staggered fermion 
actions as well as the clover \cite{clover} and (truncated) perfect actions 
for Wilson fermions \cite{Hasenfratz,Bietenholz}.

The reduction of cut-off effects plays a particularly important
role in studies of QCD thermodynamics where the relevant observables,
like energy density or pressure, are dimension four operators and the
numerical signal for these observables thus drops like the lattice spacing to
the fourth power. This forces one to perform calculations on rather
coarse lattices on which discretization errors can be significant.
In fact, this has been observed early on in studies of the bulk
thermodynamics of SU(3) gauge theories \cite{Boyd} and the efficiency
of improved actions for studies of gauge theories at high temperature has 
been demonstrated \cite{Beinlich}. 

In calculations on lattices with finite temporal extent $N_\tau$, as 
required in thermodynamic studies, cut-off effects obscure strongly 
the approach to the infinite temperature, ideal gas limit. Improvement
schemes for thermodynamic calculations therefore have been developed and
tested quite successfully in this limit for both the gauge as well as the
fermion sector of QCD. The qualitative features of the cut-off dependence 
of thermodynamic observables present in this limiting case have been found to
carry over to numerical calculations at finite temperatures and are significant
even at temperatures as low as a few times the transition temperature to the 
quark-gluon plasma phase.  

In this paper we will concentrate on a discussion of cut-off effects in 
the fermionic sector of bulk thermodynamic quantities that are given 
by derivatives of the QCD partition function with respect to
temperature, $T$, or quark chemical potential, $\mu$. In particular the
latter aspect has so far not been analyzed systematically. Some results
for the $\mu$-dependence of cut-off effects have been obtained for the
Naik and p4 staggered fermion actions \cite{Allton,milc}, truncated
perfect actions \cite{perfectmu} and for overlap
fermions \cite{Gattringer}. In fact, in the latter case the question was
raised whether the introduction of a non-zero chemical potential in the
overlap formalism could spoil the leading order cut-off dependence of
thermodynamic quantities and could reintroduce divergences of the type 
discussed in early studies of QCD at finite density \cite{Hasenfratzmu}.
We proof here that this is not the case and give explicit results for
the $\mu$-dependence of cut-off effects in leading and next-to-leading
order in a large-$N_\tau$ expansion of bulk thermodynamics. 
We show that a non-vanishing chemical potential 
only modifies the expansion coefficients but does not change the structure
of the expansion, {\it i.e.} actions that are improved to ${\cal O}(a^n)$
at $\mu=0$ remain improved to that order also for $\mu > 0$. 
We will clarify 
the relation between the improvement of 
dispersion relations and the improvement of bulk thermodynamic observables
and give explicit results for the large-$N_\tau$ expansion of the pressure
calculated with staggered and Wilson type actions. We furthermore
show that these considerations carry over to chirally invariant
fermion formulations such as overlap and domain wall fermions.

We will analyze the relation between rotational invariance of quark
propagators, improved dispersion relations and
bulk thermodynamic observables in the next two sections.
We further exemplify these results for naive staggered fermions 
as well as two popular improved staggered 
fermion actions, the Naik \cite{Naik} and the 
p4 action \cite{p4action} in section \ref{se:stagg}.
In section \ref{se:wilson} we consider fermion discretizations
of the Wilson type with general couplings on a hypercube of
size $(2a)^4$. 
We finally comment on thermodynamics with overlap and domain wall 
fermions in section \ref{se:chiral} and give our conclusions in 
section \ref{se:conclusion}.  Appendices A through D contain some
useful formulas to make the paper self-contained.

\section{Dispersion relations and the free energy}
\label{se:general}

We will start our discussion of cut-off effects for free fermions
which corresponds to the asymptotic infinite temperature limit
of QCD by considering the simplest (naive) lattice 
discretization scheme used for the fermion sector of QCD.
This will already demonstrate the main features of our analysis.
For simplicity we also will suppress color and flavor factors, 
which are just multiplicative in the infinite temperature limit.

The partition function for free fermions at temperature $T$ and for 
non-vanishing chemical potential $\mu$, defined on a lattice of size 
$N_\sigma^3 N_\tau$, is given by 
\begin{equation}
Z (V,T,\mu,m) = \int \prod_{x}{\rm d}\bar\psi_x{\rm d}\psi_x \; 
{\rm e}^{-S_F} \; ,
\label{partition}
\end{equation}
where 
$(N_\sigma a)^3 =V$ and $N_\tau a=1/T$ are the volume and inverse
temperature of the system; $a$ denotes the lattice spacing and
$S_F$ is  the Euclidean action for free fermions of mass $m$. 
 
In the simplest discretization scheme for fermionic actions derivatives
are replaced by nearest neighbor differences on a four 
dimensional lattice,
\begin{equation}
S_F = \sum_{x}\frac{1}{2}
\left( \sum_{k=1}^3 (\bar{\psi}_{x} \gamma_k \psi_{x+\hat{k}}  -
\bar{\psi}_{x}  \gamma_k \psi_{x-\hat{k}})
+{\rm e}^{\mu a}  \bar{\psi}_x \gamma_4 \psi_{x+\hat{4}}
-{\rm e}^{- \mu a} \bar{\psi}_{x}  \gamma_4 \psi_{x-\hat{4}} \right)
+ma  \bar{\psi}_x \psi_{x}  \; .
\label{naive}
\end{equation}
Here we also introduced the chemical potential  
through the usual exponential form \cite{Hasenfratzmu}.
The Grassmann-valued fermion fields obey anti-periodic
boundary conditions in the time direction. We note that the transformation
$\psi_{\vec{x}, x_4} \rightarrow {\rm e}^{-\mu x_4} \psi_{\vec{x}, x_4}$, 
$\bar{\psi}_{\vec{x}, x_4} \rightarrow {\rm e}^{\mu x_4} 
\bar{\psi}_{\vec{x}, x_4}$ leaves the path integral over the fermion fields
invariant and shifts the entire $\mu$-dependence of the fermion action
into the last time slice.
The Euclidean action thus only depends on the chemical
potential in units of the temperature, $\mu/T= \mu a N_\tau$. 
The action can be written in momentum space as 
\footnote{Here and in the following we consider massless fermions,
with obvious generalization for the massive case.}
\beq
S_F  =\sum_{p,\nu} \bar{\psi}(p) i\gamma_\nu D_\nu (p, \mu ) \psi(p)
\label{naivemomentum}
\eeq
This leads immediately to the denominator of the fermion propagator 
\beq
D(\vec{p},p_4, \mu ) \equiv \sum_{\nu=1}^4  D_\nu(\vec{p},p_4, \mu ) 
D_\nu(\vec{p},p_4, \mu )
= \sum_{k=1}^3 \sin^2(a p_k) + \sin^2(a p_4-i \mu a) \; ,
\label{eq:generalD}
\eeq
and allows to evaluate the partition function explicitly,
\begin{equation}
Z(V,T,\mu,0) = \prod_{p} D^2(\vec{p},p_4,\mu ) \; .
\label{partition_p}
\end{equation}
Here the momenta take on discrete values, $a p_k = 2\pi n_k/N_\sigma$, 
$n_k=0,~,\pm 1,...,\pm (N_\sigma/2 -1),~N_\sigma/2$ and 
$a p_4 = 2\pi (n_4+1/2)/N_\tau$, 
$n_4=0,~,\pm 1,...,\pm (N_\tau/2 -1),~-N_\tau/2$.

For our general discussion of cut-off effects that arise from the finite
lattice spacing given in units of the temperature, $aT\equiv 1/N_\tau$,
the spatial extent of the lattice is of less interest. We thus take
the thermodynamic limit, $N_\sigma\rightarrow \infty$, which simplifies
the following considerations. We then obtain the pressure of a free
fermion gas at temperature $T$ and for chemical potential $\mu$,
\begin{eqnarray}
Pa^4 &=& 2 \int_{-\pi/2}^{\pi/2} \frac{{\rm d}^3 ap}{(2 \pi)^3} 
\frac{1}{N_\tau}\sum_{0\le ap_4 \le \pi} 
\ln D(\vec{p},p_4, \mu ) \label{exact} \\
&=& 2 \int_{-\pi/2}^{\pi/2} \frac{{\rm d}^3 ap}{(2 \pi)^3} 
\frac{1}{N_\tau}\sum_{n=0}^{N_\tau/2 -1} \ln \left[ \omega^2(\vec{p}) + 
\sin^2\left( 2\pi (n+1/2)/N_\tau-i \mu a \right)\right] \; ,
\label{freeenergy}
\end{eqnarray}
with $\omega^2(\vec{p})= \sum_{k=1}^3 \sin^{2}(a p_k)$. 
Note that we reduced the integration and summation intervals to half the 
Brillouin zone which eliminates a factor 16 arising from the 16 so-called
doublers which emerge in the naive discretization. We thus normalize
to a single quark flavor as we also will do for other discretization schemes
in the following sections.
In order to evaluate
the sum over Matsubara modes appearing in Eq.~(\ref{freeenergy}) one uses the
fact that this sum can be viewed as the sum over residues resulting 
from a contour integral over poles in the complex plane \cite{Elze}, 
\beq
\sum_n f(r_n) \rightarrow \oint \frac{dr}{2 \pi i} f(r) h(r,\mu) \; .
\eeq
The computational steps given in \cite{Elze} can easily be generalized to
non vanishing chemical potential by noting that the entire
$\mu$ dependence is contained in the function $h(r,\mu)$,
\beq
h(r,\mu) = \frac{e^{\mu a N_\tau}}{r (r^{N_\tau} + e^{\mu a N_\tau})} \; ,
\eeq
the poles of which at
\beq
r_n = \exp [i ( 2\pi (n+1/2)/N_\tau-i \mu a )]
\eeq
deliver the sum over the Matsubara frequencies via
the relation
\beq
\sin^2\left( 2\pi (n+1/2)/N_\tau-i \mu a) \right)
= - \frac{1}{4} \left( r_n -\frac{1}{r_n} \right)^2 \; .
\eeq
Carrying out the sum over the discrete set of momenta yields 
\beq
\sum_{n=0}^{N_\tau/2-1}
         \ln [\sin^{2}(2 \pi(n+1/2)/N_\tau-i \mu a) + \omega^2(\vec p)] 
= \ln \left(1+ze^{-N_\tau a E(\vec p)}\right) 
+ \ln \left(1+z^{-1} e^{-N_\tau a E(\vec p)}\right)  + \mathrm{const}.
\label{gE}
\eeq
where $z = \exp(\mu a N_\tau) = \exp(\mu /T)$ is the fugacity and $E(\vec p)$
is the dispersion relation 
which is given by the pole of the fermion propagator, {\it i.e.} $E(\vec p)$
is obtained from the zeroes of Eq.(\ref{eq:generalD}) for $\mu a =0$,
\beq
D(\vec p, E = i p_4, 0) = 0 ~~ \Leftrightarrow ~~
\omega^2(\vec{p})  -\sinh^2 (a E) =0 \quad .
\label{eq:general_disp}
\eeq
This emphasizes the particular role dispersion relations play in the 
study of bulk thermodynamics. 

Inserting Eq.~(\ref{gE}) into Eq.~(\ref{freeenergy}) and subtracting the 
zero temperature part of the pressure, 
$(Pa^4)_0\equiv \lim_{N_\tau \rightarrow \infty} Pa^4$
eliminates ultra-violet divergences and
allows to define the pressure in the same normalization as it is used in
numerical calculations on the lattice,
\begin{eqnarray}
\frac{P}{T^4} &\equiv& \left[ Pa^4 - (Pa^4)_0 \right]  N_\tau^4   
\nonumber \\
&=& \frac{N_\tau^3}{4 \pi^3} \int_{-\pi/2}^{\pi/2} {\rm d}^3a p 
 \;\left[ \ln \left( 1 + z \exp \left( - N_\tau a E(\vec{p}) \right) \right)
 + \ln \left( 1 + z^{-1} \exp \left( - N_\tau a E(\vec{p}) \right) \right)
\right] \; .
\label{pressure} 
\end{eqnarray}

Eq.~(\ref{pressure}) is the starting point for a systematic analysis of cut-off
effects in bulk thermodynamic observables that are obtained from the
logarithm of the partition function, 
$P/T^4\equiv \lim_{V\rightarrow \infty} (VT^3)^{-1} \ln Z$, in terms 
of derivatives with 
respect to $T$ or $\mu$. An expansion of $P/T^4$ around $1/N_\tau = 0$ yields 
systematic corrections to the continuum ideal gas result which are given in 
terms of even powers of $1/N_\tau$. This arises after introducing 
$y_i= p_i/T = N_\tau a  p_i$ as new integration variable in Eq.~(\ref{pressure})
and realizing that the dispersion relation is an even function in the momenta. 
We will discuss this expansion in more detail in the next section.

\section{Improvement of dispersion relations and bulk thermodynamics}

\subsection{Generalized dispersion relations and the pressure}

We will outline here the arguments that lead to the observation that
also for more general fermion actions than the one discussed in the
previous section there exists a close relation between that branch of the 
dispersion relation that survives in the continuum limit and the 
large-$N_\tau$ expansion of $P/T^4$.

Similarly to the fermion propagator derived in the previous section for
the naive discretization scheme of the fermion action one obtains results
for general fermion actions of staggered or Wilson type that also may
involve more complicated terms than the 1-link term used in the 
previous section.
The propagator for free fermions of staggered type, $D(\vec p, p_4)$, can 
generally be written as a polynomial\footnote{We give here relations for
$\mu a =0$. The chemical potential can, however, be reintroduced at any
stage through the substitution $ap_4 \rightarrow ap_4 -i \mu a$.} 
in $\sin^2(a p_4)$,
\beq
D(\vec p, p_4) = \sum_{i=0}^n d_i(\vec p) \sin^{2i}(a p_4)  \; ,
\label{eq:stagg_poly}
\eeq
where the coefficients $d_i$ depend on the spatial momentum components.
For Wilson type quarks the inverse propagator can instead be written as a 
polynomial in $\sin^2(a p_4/2)$,
\beq
D(\vec p, p_4) = \sum_{i=0}^n d_i(\vec p) \sin^{2i}(a p_4/2)  \; .
\label{eq:wil_D}
\eeq
This is due to the Wilson term which introduces a 
$\cos(a p_4) = 1 - 2 \sin^2(a p_4/2)$. 
In both cases, rewriting $D$ in terms of its $n$ roots, $\omega_i(\vec p)$,
\beq
D(\vec p, k_4) = d_n(\vec p)\prod_{i=1}^n 
         \left[ \sin^{2}(a k_4) + \omega_i^2(\vec p)\right] \; ,
\label{eq:prod}
\eeq
where $k_4 = p_4$ for staggered and $k_4 = p_4/2$ for
Wilson quarks, immediately gives the $n$ dispersion relations,
\beq
\sinh^2(a \epsilon_i(\vec p)) =
- \sin^2(a k_4) = \omega_i^2(\vec p) \; , \; i=1,..,n \; ,
\eeq
with $\epsilon_i = E_i$ for staggered 
and $\epsilon_i = E_i/2$ for Wilson type quarks.
Although $D$ is real, the roots can in general be complex.
However, at least for small momenta the root which survives the continuum limit,
$\omega_1$ to be definite, must be real in order to reproduce the
continuum dispersion relation, $E^2(\vec p) = p^2$.
When we insert the right hand side of Eq.~(\ref{eq:prod}) in the partition
function, Eq.~(\ref{partition_p}), we find that
each factor appearing in the product over the $n$ distinct roots 
contributes in the evaluation of $P/T^4$ as a term in a sum over the different 
branches, $E_i$, of the dispersion relation. Each term in this sum
has the same structure as the
integral appearing in Eq.~(\ref{pressure}). In the large-$N_\tau$ limit,
however, all branches that have a gap in the dispersion relation at $p=0$
give contributions to $P/T^4$ that are exponentially suppressed for 
large-$N_\tau$. Also for more general actions Eq.~(\ref{pressure}) 
thus is the starting point for the analysis of cut-off effects in the
large-$N_\tau$ limit. In order to analyze power-like corrections to 
bulk thermodynamics in the continuum limit it suffices to analyze the 
properties of $E(\vec{p})\equiv E_1(\vec{p})$. 

\subsection{Rotational invariance, improved dispersion relations and 
improved thermodynamics}

For both, standard staggered and standard Wilson fermions
the dispersion relation receives ${\cal O}(a^2)$ corrections,
\beq
E^2(\vec p) = p^2 + {\cal O}(a^2 p^4)  
                       + {\cal O}(a^2 \sum_{k=1}^3 p_k^4) \; ,
\eeq
with $p^2 = \vec{p}^2 = \sum_{k=1}^3 p_k^2$.
Improving the dispersion relation moves the leading 
${\cal O}(a^2)$ lattice artefacts to some higher order $n$.
For this purpose it suffices to construct an
action with rotational invariance maintained to this order.
To see how this works let us briefly discuss the dispersion
relation for standard staggered fermions
which is identical to Eq.(\ref{eq:general_disp}).
Expanding this relation for small values
of $a p_k$ and $a E$ up to terms of order $p_k^4,\; E^4$ yields
\beq
E^2 - p^2 +\frac{1}{3}\sum_{k=1}^3 a^2 p_k^4 + \frac{1}{3} a^2 E^4 =0  \; ,
\label{expansion}
\eeq
which has the solution
\beq
E^2(\vec p) = p^2 - \frac{1}{3}a^2 p^4 - \frac{1}{3}\sum_{k=1}^3 a^2 p_k^4
\quad ,
\eeq
i.e. corrections to the continuum result, $E(p)=p$, start at ${\cal O}(a^2)$.
In a similar way we may proceed for improved actions which have a
Euclidean propagator that is rotationally invariant at ${\cal O}(p^4)$,
i.e. it still may contain corrections at ${\cal O}(a^2)$, however terms 
proportional to $a^2 \sum_\mu p_\mu^4$ are absent.
The important new feature is that the low momentum expansion 
of $D$, for the branch $E_1$ surviving the continuum limit,
now has the form
\beq
(E^2_1 - p^2 ) [1 + f(a p_k, a E_1)] + {\cal O}( a^4 p_\mu^6) 
= 0 \quad ,
\label{eq:E4}
\eeq
with $f(a p_k, a E_1) = {\rm const} \; a^2 (E_1^2 - p^2)$ 
which leads to
\beq
E^2_1 = p^2 +{\cal O}( a^4 p_\mu^6) \quad .
\eeq
In general, if the propagator can be factorized, 
up to order $a^n p_\mu^{n+2}$,
into a factor $E_1^2 - p^2$,
with $E_1$ corresponding to the lowest root, 
times 1 plus some function 
$f(a p_k, a E_1)$ which starts at ${\cal O}(a^2)$
then
\beq
a^{-2} D(E_1,\vec p) = 
        (E_1^2 - p^2 + {\cal O}(a^n p_\mu^{n+2})) \, 
      [1 + f(a p_k, a E_1)]
                                           + {\cal O}(a^n p_\mu^{n+2}) 
\eeq
and the dispersion relation receives ${\cal O}(p^{n+2})$ corrections only
i.e. it is ${\cal O}(a^{n-2})$ or ${\cal O}(p^n)$ improved.

Suppose now that the dispersion relation is 
${\cal O}(a^{n-2})$ improved, 
\beq
E_1^2 = p^2 + {\cal O}(a^n p^{n+2}_\mu)  \; .
\eeq
In this case the corrections to $E_1/T$ start at ${\cal O}(N_\tau^{-n})$,
\beq
\frac{E_1}{T} = N_\tau a E_1 =
\frac{p}{T} \left[ 1 + {\cal O}\left(
\frac{1}{N_\tau} \frac{p}{T}\right)^n
            \right] = 
\frac{p}{T} \left[ 1 + \Delta \right] \; ,
\eeq
where $\Delta$ is a polynomial with only even powers in $1/N_\tau$.

We may now use this expansion to evaluate corrections to the continuum
result for the pressure. To do so we rearrange
the argument of the logarithms in Eq.~(\ref{pressure});
with $y_i= p_i/T = N_\tau ap_i$, $y = p/T=\sqrt{y_1^2+y_2^2+y_3^2}$ and 
$\Delta\equiv \Delta(y_1/N_\tau,y_2/N_\tau,y_3/N_\tau)$ one has 
\beq
1 + A e^{-E_1/T} = \left( 1 + A e^{-y} \right)
\left(1 - B 
\left( 1 - e^{-y \Delta} \right) \right)  \; ,
\label{rearrange}
\eeq
where $B$ is given as $B(y,A)=1 / (  A^{-1} e^{y} +1 )$ and $A= z$ or 
$z^{-1}$, respectively. 
After expanding in $\Delta$ one then finds for the logarithms appearing
in Eq.~(\ref{pressure})
\beq
\ln \left( 1 + A \exp \left( - N_\tau a E(\vec{p}) \right) \right)
= \ln \left( 1 + A {\rm e}^ {- y}\right)  - B y \Delta +
{\cal O}(\Delta^2)  \; .
\label{lnexpand}
\eeq
While the first term will give the continuum ideal gas result for the 
pressure of a fermion gas, the second term and further higher order 
corrections will generate a systematic expansion in powers of $1/N_\tau$.
This shows that improving the dispersion relation to a certain order in
$1/N_{\tau}$ immediately leads to an improvement of the high temperature limit 
of free energy, pressure and other thermodynamic observables derived from
these to the same order. This also is 
the case for derivatives of $P/T^4$ with respect to the chemical potential, 
such as the quark number density or higher order susceptibilities.  
We will expand on this in the next subsection. In the following
it also will be convenient to introduce polar coordinates for $y_i$, which
allows us to think of the cut-off dependence entering only in the
radial component, i.e. we take the arguments of $\Delta$ to be $y/N_\tau,~
\phi,~\theta$ and rewrite it as 

\begin{equation}
\Delta \equiv \Delta (y/N_\tau , \phi,\theta) = 
\sum_{k=1}^{\infty} a_{2k}(\phi,\theta) \left( \frac{y}{N_\tau} \right)^{2k} 
\; . 
\label{polar}
\end{equation}
The lowest non-vanishing contribution will appear at $k = n/2$.

\subsection{Cut-off dependence of the pressure}

We are now in a position to discuss the systematic expansion of the 
pressure in inverse powers of $N_\tau$ and analyze systematically cut-off
effects and there dependence on temperature ($aT=1/N_\tau$) as well as
chemical potential ($\mu/T= \mu a N_\tau$). We insert Eq.~(\ref{rearrange})
into Eq.~(\ref{pressure}) and obtain
\begin{eqnarray}
\frac{P}{T^4} =&& \left(\frac{P}{T^4}\right)_{SB}    \nonumber \\
&&+ \frac{2}{(2\pi)^3} 
\int_0^{\pi N_\tau} y^2 {\rm d} y 
\int_0^{\pi} \sin \theta {\rm d} \theta 
\int_0^{2 \pi} {\rm d} \phi 
 \;\left[ \ln \left( 1 - B(y,z)\left( 1 -{\rm e}^{-y \Delta}\right)\right) 
 + \ln \left( 1 - B(y,z^{-1})\left( 1 -{\rm e}^{-y \Delta}\right)\right) 
\right] \; ,
\label{pressureNt} 
\end{eqnarray}
where the leading order ideal gas term is given by 
\begin{eqnarray}
\left(\frac{P}{T^4}\right)_{SB} &=& 
\frac{1}{\pi^2} \int_0^{\infty} {\rm d} y  y^2
 \;\left[ \ln \left( 1 + z\; {\rm e}^{- y} \right) 
 + \ln \left( 1 + z^{-1} {\rm e}^{- y}  \right)
\right] \nonumber \\
&=& \frac{4 \pi^2}{3} B_4\left(\frac{1}{2}\left(1-i \frac{\mu}{\pi
T}\right) \right) =
\frac{7\pi^2}{180}\left[1 + \frac{30}{7}\left( \frac{\mu}{\pi T}\right)^2
+\frac{15}{7} \left( \frac{\mu}{\pi T}\right)^4 \right]  \; . 
\label{pressureSB} 
\end{eqnarray}
Note that the
well-known ideal gas result for the pressure is 
an even Bernoulli polynomial $B_4$ of degree 4 in the chemical potential. 
In the following we will argue 
that similarly to the leading term also the corrections at 
order $N_\tau^{-2n}$ are Bernoulli polynomials in $\mu/(\pi T)$ of 
degree $(4+2n)$ which might be less obvious. 

As the entire cut-off dependence in 
Eq.~(\ref{pressureNt}) arises from the cut-off dependence
of the dispersion relation, {\it i.e.} from 
$\Delta = \Delta(y/N_\tau, \phi, \theta)$,
we may first expand the pressure in a Taylor series around 
$y\Delta (0,\phi,\theta)=0$. After noting that derivatives with respect to
$y\Delta$ can be replaced by derivatives with respect to $y$ we 
arrive at
\begin{eqnarray}
\frac{P}{T^4} - \left(\frac{P}{T^4}\right)_{SB} 
&=& \frac{1}{4\pi^3} \sum_{n=1}^{\infty} \frac{1}{n!} 
\int_0^{\pi N_\tau} y^2 {\rm d} y 
\int_0^{\pi} \sin \theta {\rm d} \theta 
\int_0^{2 \pi} {\rm d} \phi 
(y \Delta)^n 
\left( \frac{\partial^{n-1}B(y,z)}{\partial y^{n-1}} +
\frac{\partial^{n-1}B(y,z^{-1})}{\partial y^{n-1}}\right)  \; . 
\label{pressureTaylor} 
\end{eqnarray}
Using the representation of $\Delta$ given in Eq.~(\ref{polar})
we now can rearrange the above series and obtain an expansion
in even powers of $1/N_\tau$, 
\begin{eqnarray}
\frac{P}{T^4} - \left(\frac{P}{T^4}\right)_{SB} 
&=& \sum_{k=1}^{\infty} A_{2k}  
P_{2k}\left( \frac{\mu}{\pi T} \right)
\left( \frac{\pi}{N_\tau}\right)^{2k} 
\; ,
\label{pressureNtk} 
\end{eqnarray}
where $A_{2k}$ results from the integration over angular variables
and the chemical potential dependent part, $P_{2k}\left( \mu/\pi T \right) $,
is given in terms of a Bernoulli polynomial that results from the integration
over the radial coordinate $y$, 
\begin{eqnarray}
P_{2k}\left(\frac{\mu}{\pi T}\right) &\equiv&
\frac{1}{(2^{-3-2k}-1)\; b_{4+2k}}
B_{4+2k}\left( \frac{1}{2} - i \frac{1}{2} \frac{\mu}{\pi T} \right) 
\nonumber \\
&=&
\sum_{l=0}^{k+2} (-1)^{2+k-l} { 4+2k \choose 2l } 
\frac{2^{2l-1}-1}{2^{3+2k}-1} \frac{b_{2l}}{b_{4+2k}} 
\left( \frac{\mu}{\pi T}\right)^{4+2k-2l}
\label{Bernoulli}
\end{eqnarray}
with $b_{n}$ denoting Bernoulli numbers.
The functions $P_n$ have been 
normalized such that $P_n(0)=1$. 
Note that the ideal
gas result can similarly be written as
$A_0 P_0(\mu/\pi T)$ with $A_0 = 7 \pi^2 / 180$. 
We give some more details on the derivation
of Eq.~(\ref{pressureNtk}) and the definition of $A_{2k}$ in Appendix~\ref{app:pressure}. 
The point we want to stress here is that the dependence
of cut-off effects on the chemical potential arises only through a 
Bernoulli polynomial of degree $4+2k$.
For the particular complex argument appearing in Eq.~(\ref{Bernoulli})
the Bernoulli polynomial contains only even powers of the chemical potential,
reflecting the particle anti-particle symmetry, $\mu \leftrightarrow -\mu$,
of the partition functions. We also note that all terms in Eq.~(\ref{Bernoulli})
are positive. Cut-off effects thus become larger with increasing $\mu/T$, although
the effect is small as exact evaluations of the pressure for fixed $N_\tau$
show \cite{Allton}.

Eqs.~(\ref{pressureNtk})  and (\ref{Bernoulli}) are the main result of this 
section. With this we have
shown that in the high temperature, ideal gas limit of QCD cut-off effects 
in the fermion sector of bulk thermodynamic observables can be traced back 
to the cut-off dependence of the dispersion relation for the quark
propagator. The structure of the cut-off dependence is preserved at 
non-zero chemical potential $\mu$, {\it i.e.} actions which are improved 
to a certain order in $aT\equiv 1/N_\tau$ at $\mu=0$ are so also for 
$\mu > 0$. Moreover, we find that the $\mu$-dependence 
of cut-off effects is universal, {\it i.e.} $\mu$-dependent correction factors
at ${\cal O}(N_\tau^n)$ are independent of the discretization and improvement
schemes and are proportional to a Bernoulli polynomial of degree $4+n$.

In the following we give explicit results for cut-off dependent corrections
to the continuum ideal gas limit for several fermion actions and compare
the asymptotic large $N_\tau$ behavior derived in this section with 
an explicit evaluation of partition functions at small values of $N_\tau$.  

\section{Staggered type quarks}
\label{se:stagg}

We now want to apply the general considerations presented in the previous
section to the case of staggered fermions. We restrict ourselves to the
class of staggered actions which contain terms with fermion and anti-fermion
fields separated by up to three links, 
\beq
S_F = \sum_{x,y} \bar{\chi}(x)  M(x,y) \chi(y)  \; ,
\eeq
with
\bew
M(x,y) & = & \sum_{\mu=1}^4 \eta_\mu (x) \left( \sum_{i=1,3} c_{i,0}
           \left[ \delta(x+i \hat \mu,y) - \delta(x-i \hat \mu,y) \right]
           \right.
           \nonumber \\
 &   & + \left. 
             \sum_{\nu \neq \mu} \sum_{j= \pm 2} c_{1,j}
           \left[ \delta(x+ \hat \mu + j \hat \nu,y) 
                - \delta(x- \hat \mu + j \hat \nu,y) \right]
           \right)   \; .
\eew
To reproduce the correct continuum limit the expansion coefficients are
constrained by the relation
\begin{eqnarray}
& &c_{1,0} + ~3 c_{3,0} + 6 c_{1,2} = {1 \over 2} \; .
\label{continuumconstraint}
\end{eqnarray}
Within this class of staggered fermion actions one can construct actions 
that are rotationally invariant up to ${\cal O} (p^4)$, i.e. the 
denominator of the fermion propagator,
$D(\vec{p})$, becomes a function of only $p^2$ up to this order.
This can be achieved with the additional constraint \cite{p4action}
\begin{eqnarray}
& &c_{1,0} + 27 c_{3,0} + 6 c_{1,2} = 24 c_{1,2} \; \;
\xrightarrow{Eq.(\protect\ref{continuumconstraint})} \;\; c_{1,2}=1/48 + c_{3,0} \; .
\label{threelink}
\end{eqnarray}
Particularly well-known versions of 3-link actions that are currently
used in studies of QCD thermodynamics with staggered fermions are the
Naik action \cite{Naik}, for which $c_{1,2}\equiv 0$, and the p4 action \cite{p4action} 
where $c_{3,0}\equiv 0$.
 
For the general three link action, not yet restricted to the subset of 
rotational invariant propagators, the propagator in momentum space 
is a polynomial of degree 3 in $\sin^2(a p_4)$,
{\it i.e.} it is of the generic form given in Eq.(\ref{eq:stagg_poly}). 
Following the discussion presented in the previous section one may
work out the dispersion relation on the branch that survives in the continuum
limit. For the leading correction to the continuum result one finds 
\begin{equation}
\frac{E_1}{p}  = 1 -\frac{1}{6} (1-48(c_{1,2}-c_{3,0})) \left( p^2 +
\frac{1}{p^2}\sum_{k=1}^3 p_k^4 \right) a^2 + {\cal O}(a^4) \; ,
\label{dispersion_naive}
\end{equation}
where we have used Eq.~(\ref{continuumconstraint}) to eliminate $c_{1,0}$.

As corrections to the continuum dispersion relation, $E_1=p$, start at
${\cal O}(a^2)$, bulk thermodynamic observables will receive cut-off
dependent corrections starting at the same order, $(aT)^2=1/N_\tau^2$. 
Eqs.~(\ref{dispersion_naive}) and (\ref{threelink}), however, 
show that all three link 
actions with a rotationally invariant propagator will lead to an
${\cal O}(a^2)$ improved  dispersion relation and thus also to 
${\cal O}(a^2)$ improved bulk thermodynamics. 
In the class of three link staggered fermion actions one finds for the
${\cal O}(a^2)$ improved dispersion relations,

\begin{eqnarray}
\frac{E_1}{p}  = &&1  
+ \frac{3}{40} \left\{ p^4 - \frac{1}{p^2}\sum_{k=1}^3 p_k^6 \right\} a^4 \nonumber \\ 
&&- \frac{1}{756} (1+1344 c_{3,0})) \left( p^3 -
\frac{1}{p}\sum_{i<j} p_i^2 p_j^2 \right)^2 a^6 +
\frac{1}{14} (p_1 p_2 p_3)^2 a^6 
+ {\cal O}(a^8) \; .
\label{dispersion_improved}
\end{eqnarray}  
We note that all three link actions lead to identical leading order
corrections to the continuum dispersion relation, but differ at 
subleading order. According to the discussion presented in the previous
section, this feature will carry over also to the cut-off dependence
of bulk thermodynamics at zero and non-zero chemical potential. 

We have evaluated the leading cut-off dependent corrections to the 
continuum ideal gas pressure for a generic 3-link, staggered fermion action
up to ${\cal O}(1/N_\tau^6)$ using Mathematica to perform the expansion
of the integrand appearing in Eq.~(\ref{pressureNt}) in inverse powers of 
$N_\tau$.  For the pressure we find,
\begin{eqnarray}
\frac{180}{7\pi^2}\left[ \frac{P}{T^4} - \left(\frac{P}{T^4}\right)_{SB}
\right] 
&=& 
\frac{248 (1- 48 (c_{1,2}-c_{3,0}))}{147} P_2\left(\frac{\mu}{\pi T}\right)
\left( \frac{\pi}{N_\tau}\right)^2 \nonumber \\
&&+\left( \frac{635}{147} +\frac{16256 (c_{3,0}-c_{1,2})}{35}
+\frac{2365248 (c_{3,0}-c_{1,2})^2}{245} \right) P_4\left(\frac{\mu}{\pi T}\right) 
\left( \frac{\pi}{N_\tau}\right)^4 \nonumber \\
&&+ \frac{73}{2079} \left( 1 + 6528 c_{3,0}
+(1-48 (c_{1,2}-c_{3,0})) (571 + 1018368 c_{1,2}^2  \right.\nonumber\\
&&\left. ~~~+51888 c_{3,0} + 
1331712 c_{3,0}^2 -48 c_{1,2} (1217 + 48960 c_{3,0})) \right)
P_{6}\left(\frac{\mu}{\pi T}\right) 
\left( \frac{\pi}{N_\tau} \right)^6 \nonumber \\
&&+ {\cal O}((\pi/N_\tau)^8) \; ,
\label{staggered_3link}
\end{eqnarray}
with $P_n$ defined in Eq.~(\ref{Bernoulli}). Results for the 
standard staggered (1-link) action as  well as for the 
class of ${\cal O}(a^2)$ improved 3-link actions, for which 
$c_{1,2}=1/48+c_{3,0}$,
are easily read off from Eq.~(\ref{staggered_3link}). In 
Table~\ref{tab:staggered} we give the expansion coefficients 
for the standard staggered action as well as the Naik and p4 actions.
We note that coefficients of the sub-leading corrections to the
pressure calculated with the standard staggered action as well as with the Naik action 
are larger than the leading expansion coefficient. The subleading correction for
the Naik action is a factor $137$ larger than for the p4-action.

\begin{table}
\begin{tabular}{|c|c|c|c|}
\hline
action & $A_{2}/A_{0}$ & $A_{4}/A_{0}$ & $A_{6}/A_{0}$\tabularnewline
\hline
\hline
standard staggered & $248/147$ & $635/147$ & $3796/189$\tabularnewline
\hline
Naik & $0$ & $-1143/980$ & $-365/77$\tabularnewline
\hline
p4 & $0$ & $-1143/980$ & $73/2079$\tabularnewline
\hline
\hline
standard Wilson & $248/147$ & $635/147$ & $13351/8316$\tabularnewline
\hline
hypercube & $-0.242381$ & $0.114366$ & $-0.0436614$\tabularnewline
\hline
\hline
overlap/ & $248/147$ & $635/147$ & $3796/189$\tabularnewline
domain wall &~&~&\tabularnewline
\hline
\end{tabular}
\caption{Coefficients in the series expansion (Eq.~(\ref{pressureNtk})) 
for the pressure calculated using different actions. The expansion
coefficients have been normalized to the ideal Fermi gas value
at vanishing chemical potential, 
$A_0\equiv(p/T^4)_{SB}(\mu/T=0) = 7\pi^2/180$.
}
\protect\label{tab:staggered}
\end{table}

\begin{figure}
\epsfig{file=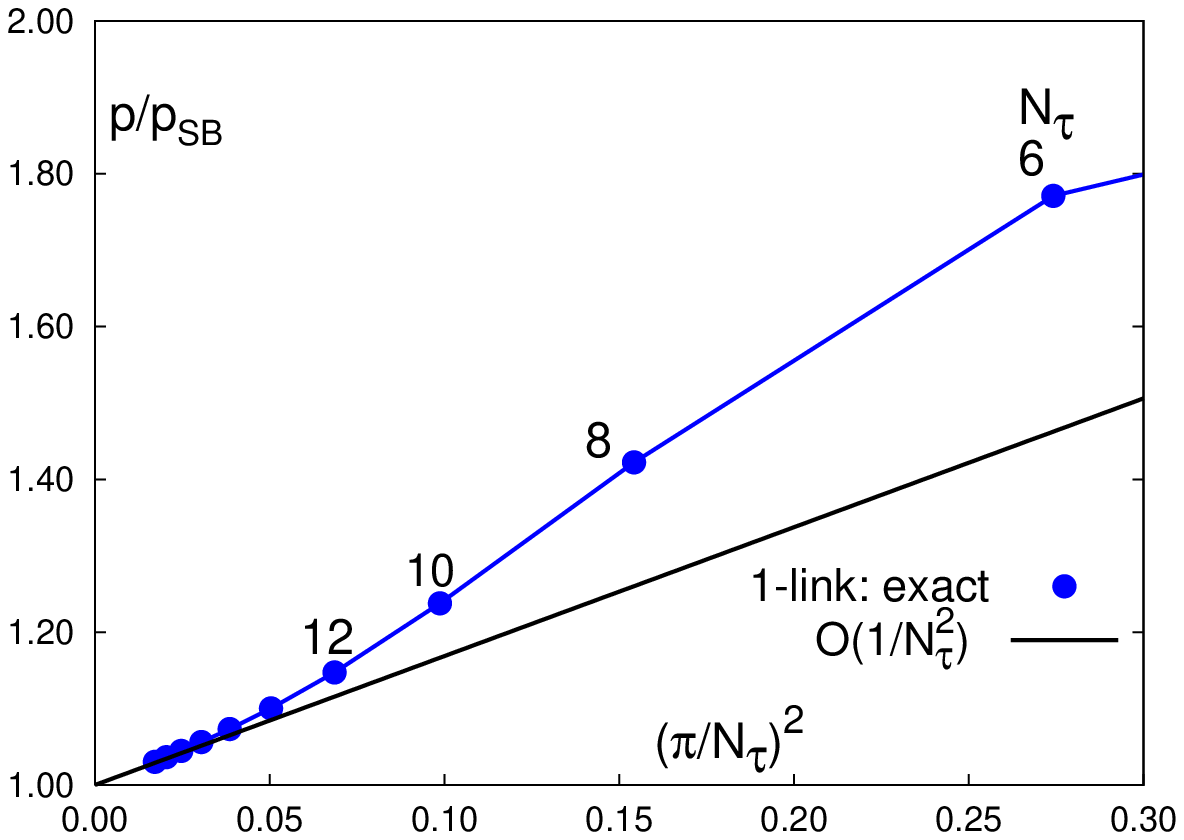,width=7.5cm}
\epsfig{file=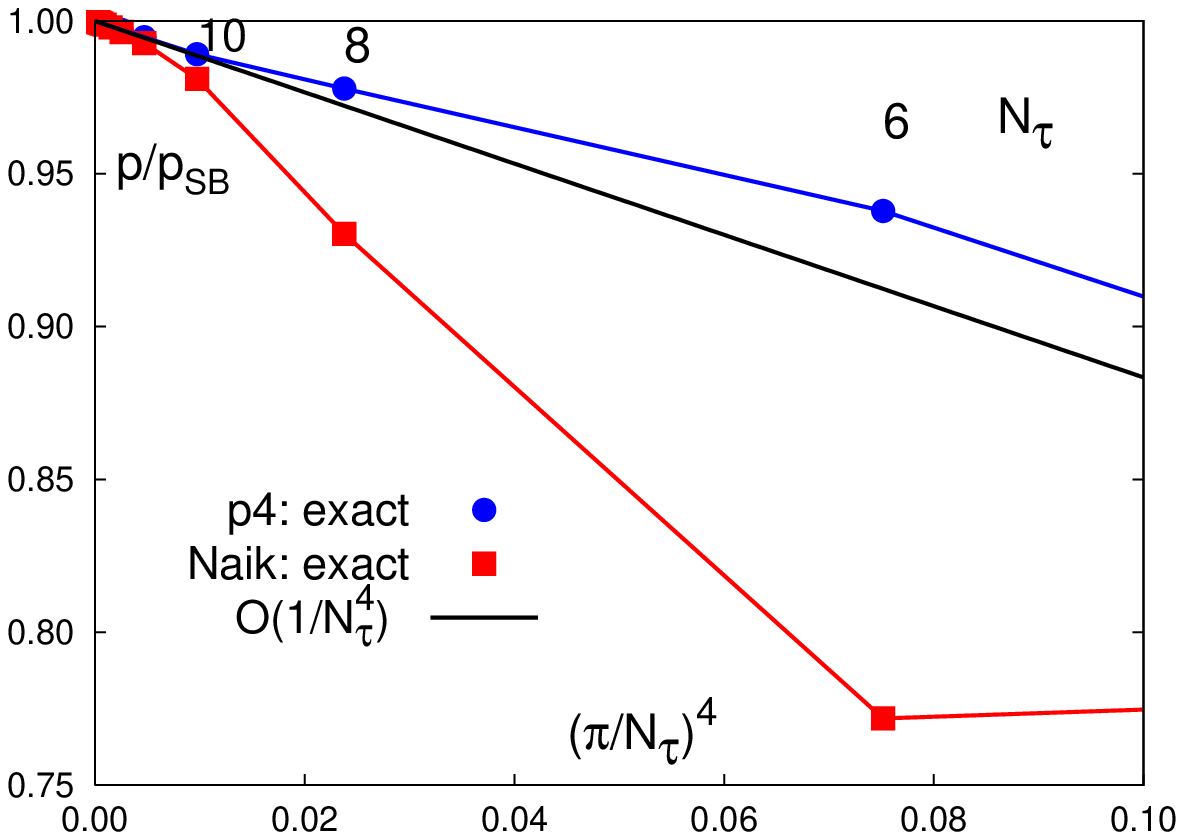,width=7.5cm}
\caption{The pressure for staggered type quarks
normalized to the continuum Stefan Boltzmann limit. The left hand
figure shows results for the (standard) 1-link action and the left hand figure
gives results for the Naik and p4 actions. In both cases we compare
exact results, calculated on lattices with temporal extent $N_\tau$
and infinite spatial extent, with the leading $1/N_\tau^n$ correction.
Note the different vertical scales.}
\label{fig:stagg_f0}
\end{figure}

It is instructive to compare the asymptotic large $N_\tau$ behavior
of the cut-off dependent corrections to the continuum ideal Fermi gas
with the exact results for the pressure evaluated at fixed values of
$N_\tau$ by using Eq.~(\ref{exact}). At vanishing chemical potential this is 
shown for standard 
staggered, Naik and p4 improved fermions in Fig.\ref{fig:stagg_f0}. 
Here we have normalized the exact lattice results to the continuum 
Stefan-Boltzmann limit, $ 7 \pi^2 / 180$. We note that the exact
calculation of the partition function at fixed values of $N_\tau$ 
reproduces the basic pattern derived from the leading orders of the
large-$N_\tau$ expansion. Cut-off effects in the pressure are reduced 
when using improved discretization schemes.

\section{Wilson type quarks}
\label{se:wilson}

We will now consider the case of Wilson fermions. Since in the interacting
case cut-off effects arise in the Wilson fermion formulation already 
at ${\cal O}(ag^2)$
improvement schemes developed for Wilson fermions have primarily focused
on  removing terms linear in the cut-off (clover improved 
actions \cite{clover}). 
Actions that are ${\cal O} (a^2 g^0)$ improved received much less 
attention and, in fact, are currently not exploited in studies of 
QCD thermodynamics with Wilson fermions. Nonetheless, removing or at least
reducing ${\cal O} (a^2)$ effects in the Wilson fermion formulations is
as important as in the staggered case when one wants to perform 
studies of QCD thermodynamics at high temperature within this 
discretization scheme. We will discuss here a generic Wilson type action 
with couplings constrained to a hypercube of size $(2a)^4$. 
This type of action is discussed in the context of truncated fixed point 
actions \cite{Bietenholz,perfectnew}. Some exploratory studies of
thermodynamics with dynamical hypercube fermions have been presented in
\cite{Shcheredin}. The Dirac matrix for this generic 
Wilson action can be written as
\beq
M(x,y) = \sum_{\mu=1}^4 \gamma_\mu \rho_\mu (x-y) + \lambda(x-y)
\eeq
with
\bew
\rho_\mu(x-y) & = & 
      ~~ \rho_1 ~~~ [ \delta(y,x+\muh) - \delta(y,x-\muh) ] \nonumber \\
&&  + \rho_2 \sum_{\nuh}
           [ \delta(y,x+\muh+\nuh) - \delta(y,x-\muh+\nuh) ] \nonumber \\
&&  + \rho_3 \sum_{\nuh , \rhoh}
           [ \delta(y,x+\muh+\nuh+\rhoh) - \delta(y,x-\muh+\nuh+\rhoh) ]
                                                       \nonumber \\
&&  + \rho_4 \sum_{\nuh , \rhoh, \sigmah}
           [ \delta(y,x+\muh+\nuh+\rhoh+\sigmah) - 
                      \delta(y,x-\muh+\nuh+\rhoh+\sigmah) ]
\eew
for the vector terms and
\bew
\lambda (x-y) & = & ~~ \lambda_0 ~~ \delta(y,x) \nonumber \\
&&  + \lambda_1 \sum_{\muh} [ \delta(y,x+\muh) + \delta(y,x-\muh) ] \nonumber \\
&&  + \lambda_2 \sum_{\muh, \nuh}
           [ \delta(y,x+\muh+\nuh) + \delta(y,x-\muh+\nuh) ] \nonumber \\
&&  + \lambda_3 \sum_{\muh, \nuh , \rhoh}
           [ \delta(y,x+\muh+\nuh+\rhoh) + \delta(y,x-\muh+\nuh+\rhoh) ]
                                                       \nonumber \\
&&  + \lambda_4 \sum_{\muh, \nuh , \rhoh, \sigmah}
           [ \delta(y,x+\muh+\nuh+\rhoh+\sigmah) + 
                      \delta(y,x-\muh+\nuh+\rhoh+\sigmah) ]
\eew
for the scalar ones. The sums over $\nuh, \rhoh, \sigmah$
extend over positive and negative directions
and are mutually orthogonal to each other and to $\muh$.
Examples for this type of action are the standard Wilson action (including
the clover improved version of it) and the hypercube truncated perfect
action \cite{Bietenholz,perfectnew}, with coefficients
as listed in Table \ref{tab:coeff} for the massless case. 

\begin{table}[t]
\begin{center}
\begin{tabular}{|c|r|c||c|r|c|}
\hline
       & Hypercube & Wilson &
       & Hypercube & Wilson \\
\hline
\hline
            &              &      &
$\lambda_0$ &  1.852720547 & 4    \\
$\rho_1$    &  0.136846794 & 1/2  &
$\lambda_1$ & -0.060757866 & -1/2 \\
$\rho_2$    &  0.032077284 & 0    &
$\lambda_2$ & -0.030036032 & 0    \\
$\rho_3$    &  0.011058131 & 0    &
$\lambda_3$ & -0.015967620 & 0    \\
$\rho_4$    &  0.004748991 & 0    &
$\lambda_4$ & -0.008426812 & 0    \\
\hline
\end{tabular}
\end{center}
\vspace*{-3mm}
\caption{Coefficients $\rho_i$ and $\lambda_i$ for standard
         Wilson quarks and for the hypercube action.}
\label{tab:coeff}
\end{table}

\medskip

The denominator of the fermion propagator, $D(p)$, is most conveniently 
written as
\beq
D(\vec p,p_4) =
R (\vec p) + 2 P(\vec p) \cos (a p_4) + Q(\vec p) \cos^2 (a p_4) \; .
\label{eq:Dwil}
\eeq
Note that this complies with Eq.(\ref{eq:wil_D}) since $\cos(a p_4)$ is to 
be rewritten as $1 - 2 \sin^2 (a p_4 /2)$.
The coefficients $P, Q$ and $R$ depend on $\vec p$
and are listed in Appendix~\ref{app:hypercube}. The requirement that in the limit of
$a p_\mu \rightarrow 0$ for all $\mu = 1,..,4$ the continuum dispersion
relation is obtained leads to the constraints (see also \cite{Adams})
\bs
2 \rho_1 + 12 \rho_2 + 24 \rho_3 + 16 \rho_4 & = & 1 \; , \\
\lambda_0 + 8 \lambda_1 + 24 \lambda_2 + 32 \lambda_3 + 16 \lambda_4 & = & 
0 \; .
\label{cont_W}
\es
Moreover, demanding that the so-called doublers receive a non-vanishing
mass at the corners of the Brillouin zone,
$a p_\mu = \pi$ for one or more component $\mu$,
introduces further conditions to be satisfied by the $\lambda$
coefficients,
\bs
\lambda_0 + 4 \lambda_1                - 16 \lambda_3 - 16 \lambda_4 & \neq & 0
\\
\lambda_0               -  8 \lambda_2                + 16 \lambda_4 & \neq & 0
\\
\lambda_0 - 4 \lambda_1                + 16 \lambda_3 - 16 \lambda_4 & \neq & 0
\\
\lambda_0 - 8 \lambda_1 + 24 \lambda_2 - 32 \lambda_3 + 16 \lambda_4 & \neq & 0
\es
In the standard Wilson action these conditions are realized such that the
so-called doubler masses are $2 /a, 4 /a, 6 /a$ and $8 /a$ 
while for hypercube fermions these masses are $\simeq 2 /a$ 
with deviations of at most 2 per cent for all corners \footnote{
Note that with five equations the $\lambda$ coefficients can be
determined uniquely once the doubler masses have been fixed. Indeed, fixing
these masses to $2/a$ leads to the HF scalar couplings within 2 per cent
accuracy.}.

We may now go through the steps outlined in Section II to determine 
the cut-off dependence of dispersion relations and bulk thermodynamics.
We start by expanding the quark propagator, Eq.~(\ref{eq:Dwil}), for
small momenta, taking into account the constraints given in Eq.~(\ref{cont_W}),
\begin{equation}
D(\vec{p},p_4) = (pa)^2 - \frac{1}{3}(-1 + 12 \rho_2 + 48 \rho_3 + 48 \rho_4)
\left( (pa)^4 - \sum_{i=1}^{4} (ap_i)^4 \right) + {\cal O} (a^6) \; .
\label{D_expand}
\end{equation}
With the couplings in the hypercube action as listed in 
Table~\ref{tab:coeff} the propagator receives ${\cal O}(a^2)$ corrections,
which however are small when compared to the standard 1-link Wilson action.
For the latter the prefactor in Eq.~(\ref{D_expand}) equals $1/3$ while
for the hypercube action it is $-0.0479$. Of course, with  
an appropriate choice of the couplings $\rho_2,\; \rho_3$ and $\rho_4$ one
can also eliminate the ${\cal O} (a^2)$ corrections to the 
propagator of the hypercube Wilson action completely, making it rotational 
invariant up to that order. 

Turning now to the dispersion relation, from Eq.(\ref{eq:Dwil})
one obtains, for the branch relevant to the continuum limit,
\beq
\sinh^2\left(\frac{a E_1}{2}\right) = - \frac{1}{2}
\left(
1 + \frac{P + \sqrt{P^2 - Q R}}{Q}
\right) \; .
\label{eq:wil_e}
\eeq
The low momentum expansion of this dispersion relation yields,
\beq
 \frac{E_1}{p} =  1 +  \frac{1}{6}(-1 + 12 \rho_2 + 48 \rho_3 + 48 \rho_4)
\left( p^2 + \frac{1}{p^2} \sum_{k=1}^3 p_k^4
\right) a^2 + {\cal O} (a^4)  \; .
\label{dispersion_W}
\eeq
We note that for standard Wilson fermions ($\rho_2=\rho_3=\rho_4=0$) the leading 
order correction to the continuum dispersion relation is identical to
that of staggered fermions. This will, of course, carry over also to 
the leading corrections to bulk thermodynamic observables.
Moreover, a choice $-1 + 12 \rho_2 + 48 \rho_3 + 48 \rho_4 = 0$
not only removes the rotational symmetry breaking term,
$\sim \sum_{k=1}^3 p_k^4$, in the propagator and the dispersion relation,
but eliminates the ${\cal O}(a^2)$ corrections completely.

\begin{figure}
\epsfig{file=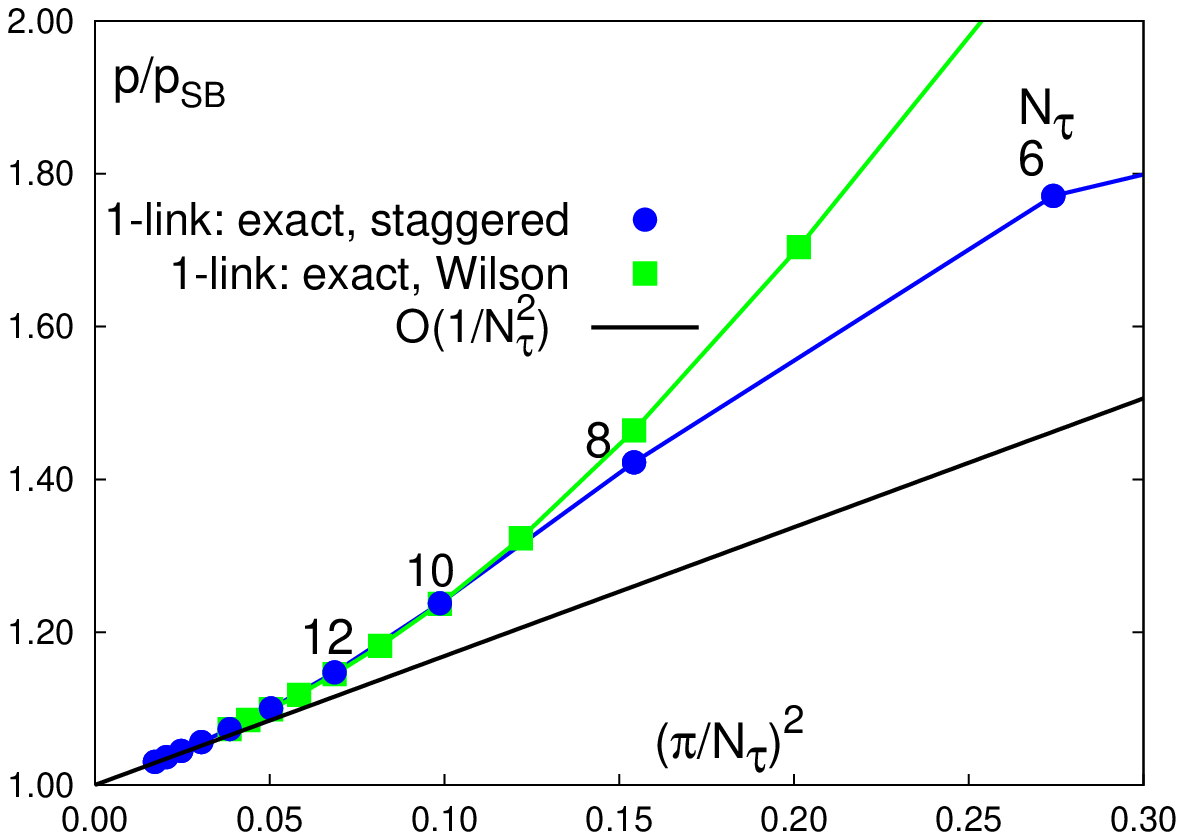,width=7.5cm}
\epsfig{file=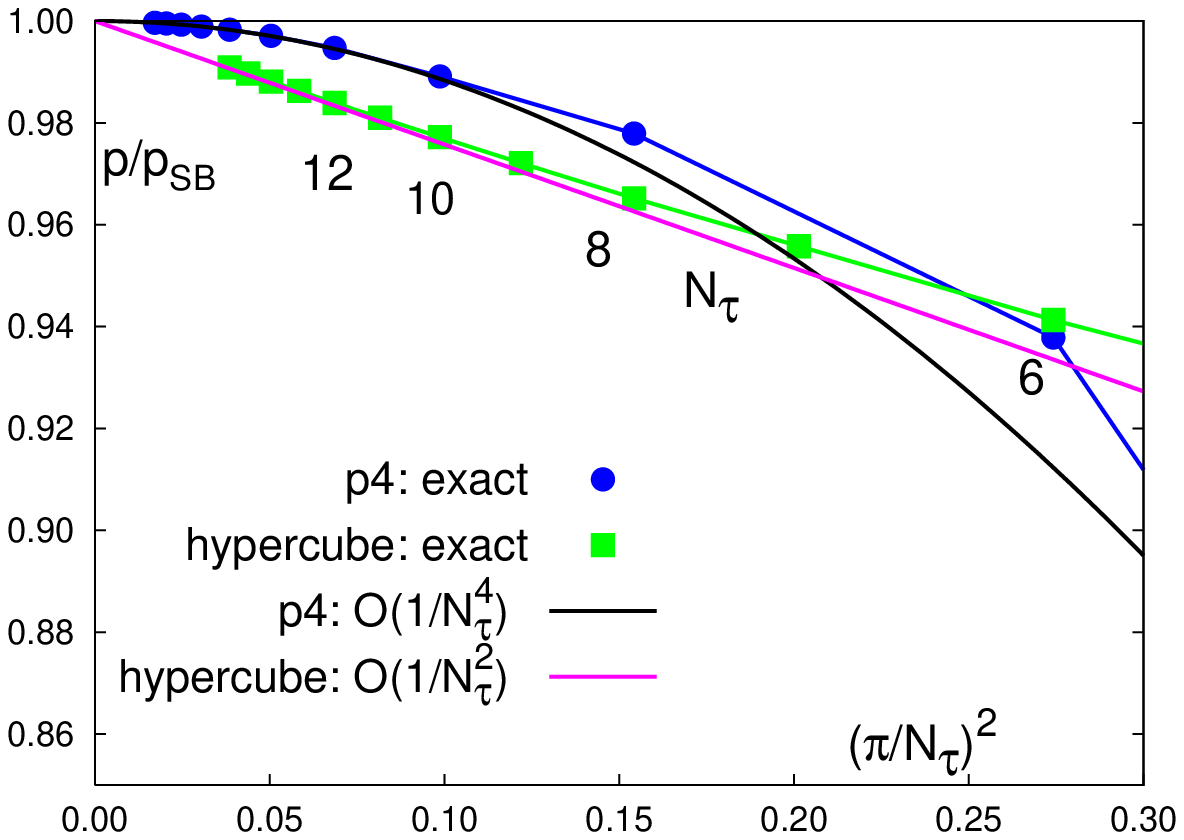,width=7.5cm}
\caption{Pressure for standard Wilson fermions (left) and 
hypercube Wilson fermions (right) normalized to the
continuum Stefan Boltzmann result. In the left hand part
we compare the results obtained with standard Wilson fermions to that
of standard staggered fermions. Both discretization schemes have 
identical cut-off errors in leading order $1/N_\tau^2$. The right hand 
part compares results for the hypercube action with the p4 action.
 }
\label{fig:f0_w}
\end{figure}

Using the dispersion relation for the hypercube action, 
Eq.~(\ref{eq:wil_e}), in our basic relation for the pressure, 
Eq.~(\ref{pressureNt}), we again can perform a systematic expansion
for large $N_\tau$. With this we obtain for the leading cut-off
corrections to the continuum pressure,
\begin{eqnarray}
\frac{180}{7\pi^2}\left[ \frac{P}{T^4} - \left(\frac{P}{T^4}\right)_{SB}
\right] 
&=& 
-\frac{248}{147} (-1 + 12 \rho_2 +48 \rho_3 +48 \rho_4) 
P_2\left(\frac{\mu}{\pi T}\right) \left(\frac{\pi}{N_\tau}\right)^2 \nonumber \\
&&+\frac{127}{735} \left( 25+3492 \rho_2^2 +55872 \rho_3^2 -2148 \rho_4 + 55872 \rho_4^2
\right. \nonumber \\ 
&&\hspace*{0.7cm}\left. 
+2328 \rho_3 (-1+48 \rho_4) + 3 \rho_2 (-209+9312 \rho_3 + 9312 \rho_4) \right)
P_4\left(\frac{\mu}{\pi T}\right) \left(\frac{\pi}{N_\tau}\right)^4 \nonumber\\
&&+h_6 P_6\left(\frac{\mu}{\pi T}\right) \left(\frac{\pi}{N_\tau}\right)^6 
+{\cal O}((\pi/N_\tau)^8) \; .
\label{Wilson_P}
\end{eqnarray}
The explicit form of $h_6$ is rather lengthy. It is given in Appendix~\ref{app:fp6}.

In Fig.~\ref{fig:f0_w} we compare exact results for the pressure obtained
with standard Wilson fermions and the hypercube action with the leading 
order large-$N_\tau$ expansion obtained in this section. As can be seen
in Fig.~\ref{fig:f0_w}(left) the standard Wilson action and standard
staggered fermions have identical leading order cut-off errors and consequently
follow an identical pattern for large $N_\tau$, e.g. for $N_\tau \gsim 10$. 
Cut-off effects for the hypercube action are significantly reduced and
of similar magnitude as for the improved staggered actions 
(Fig.~\ref{fig:f0_w}(right)). Cut-off dependent corrections at small values 
of $N_\tau$ are 
already close to the result obtained from the leading order expansion
reflecting that higher order corrections are small. This also is apparent 
from the 
higher order corrections listed in Table~\ref{tab:staggered}. 

\section{Overlap and domain wall fermions}
\label{se:chiral}

Let us now comment on bulk thermodynamics with chirally invariant fermion
formulations, e.g. overlap \cite{Neuberger} and domain wall fermions 
\cite{Shamir,Kaplan}. The free field limit of chiral fermion formulations 
\cite{Capitani} has recently been used to analyze the cut-off dependence
of spectral functions at high temperature \cite{Aarts}. We follow this 
discussion here closely to explore the cut-off dependence of bulk 
thermodynamics in lattice formulations that use chiral fermion formulations,
e.g. overlap or domain wall fermions with standard Wilson kernel.

It is known that in the free field limit the dispersion relations for
overlap and domain wall fermions are given by the standard (staggered) 
dispersion
relation up to a constraint, which only alters the large momentum part of
the dispersion relations \cite{Capitani}. As we have seen in the previous 
sections this part 
of the spectrum does not influence the large-$N_\tau$
expansion of thermodynamic quantities. Following the general discussion
of cut-off effects given in Section II we thus expect that
the $1/N_\tau$ expansion for the cut-off dependence of the pressure 
and related thermodynamic quantities is identical to that of standard
staggered fermions. As outlined in Section II this also includes the 
dependence on a non-zero chemical potential. Nonetheless, in particular
for the case of overlap fermions the question has been raised whether
the fact that one has to take a square root of e.g. the Wilson fermion operator
may lead to difficulties in reproducing the correct continuum thermodynamics
at non-zero chemical potential \cite{Gattringer}. 
Indeed, the presence of a square root in the basic relation for the 
partition function (see Eq.~(\ref{opres})) leads to cuts which so far did not
show up in our general discussion of bulk thermodynamics in the ideal
gas limit. We will, however, show that the cut contributions are
exponentially suppressed and do not contribute to the $1/N_\tau$ expansion
of the pressure.

We outline here the basic steps 
that allow to relate also in the overlap formalism the basic
properties of the dispersion relation with the cut-off dependence of
the pressure. Some more technical considerations are given in 
Appendix \ref{app:overlap}.  

We work with the following form of the (massless) overlap operator
\cite{Capitani,Aarts}, 
\beq 
D = \frac{1}{a} \left(1 + \frac{X}{\sqrt{X^{\dagger}X}}\right) \; .
\eeq
Here $X=D_W-1$, where $D_W$ is a nonchiral, undoubled fermion operator like
the standard Wilson operator discussed in the previous section. In momentum
space $X(\vec{p},p_4)$ takes on the form
\beq
X(\vec{p},p_4) = =\frac{1}{a}\left( i \sum_{\nu=1}^{4} \gamma_\nu
\sin(ap_\nu) + 2 \sum_{\nu=1}^{4} \sin^2(ap_\nu /2) -1\right) \; .
\label{Xp}
\eeq
In order to keep the notation simple we do not explicitly introduce a 
chemical potential.
It can, however, at any step be re-introduced through the
replacement $ap_4\rightarrow ap_4 - i\mu a$.

The fermion propagator of the overlap operator is then given by
\beq
D(\vec{p},p_4) = \frac{1}{2}\left( 1 - \frac{\sum_{\nu=1}^{4} i \gamma_\nu
\sin(ap_\nu)}{\sqrt{S^2+A^2} -A} \right) \; ,
\label{oprop}
\eeq
where
\begin{eqnarray}
A\equiv A(\vec{p},p_4) = 1 - 2 \sum_{\nu=1}^{4} \sin^2(ap_\nu /2) & & 
S^2\equiv S^2(\vec{p},p_4)=\sum_{\nu=1}^4 \sin^2(ap_\nu)  \; .
\label{AS}
\end{eqnarray}
From Eq.~(\ref{oprop}) we see that the poles of the fermion propagator
are given by 
\begin{eqnarray}
\sqrt{A^2+S^2} -A = 0 \quad \Leftrightarrow \quad
S^2 = 0 
\;\; \text{and} \;\;  A > 0  \; .
\end{eqnarray}
Thus, overlap fermions have indeed exactly the same dispersion relation 
as naive 
fermions upto a constraint, $A>0$. The constraint is automatically satisfied 
at small momenta but has no solutions near the edge of the Brillouin zone, 
thus avoiding the doubling problem. As pointed out in \cite{Aarts} this
carries over to similar relations for the propagator and the poles of free
domain wall fermions in the limit of infinite 5th dimension.

With this we obtain for the pressure 
\begin{eqnarray}
Pa^{4} &=& \frac{2}{(2\pi)^3} \int_{-\pi}^{\pi} {\rm d}^3 ap 
\frac{1}{N_\tau} \sum_{ap_4} \ln\left(1-\frac{A}{\sqrt{A^2+S^2}}
\right)  
\; .
\label{opres}
\end{eqnarray}
In Eq.(\ref{opres}) the appearance of a square root, of course, does not create
any problems; its argument is positive and the pressure thus can be evaluated
for any fixed temporal lattice extent $N_\tau$. The discussion of the
large-$N_\tau$ expansion, however, becomes somewhat more complicated than in
the cases considered in the previous sections. We also note that we did not 
reduce the integration range, as it has been done for naive fermions to
eliminate contributions from doublers. Integrals and momentum sums still 
extend over the whole Brillouin zone.
Let us denote $x\equiv \sin(ap_4/2)$. The expression for the pressure may be 
shown to take the form 
\beq
Pa^{4} = \frac{2}{(2\pi)^3} \int_{-\pi}^{\pi} {\rm d}^3 ap 
\frac{1}{N_\tau} \sum_{ap_4}
\ln\left(1+\frac{1}{\sqrt{2}}\frac{x^2+c^2-1/2}{\sqrt{c^2x^2+b^2}}\right) \; ,
\eeq
where \begin{eqnarray}
b^2 &=& \sum_{i<j}\sin^2 ap_i/2\sin^2 ap_j/2 + 1/8 = 
\frac{1}{8}\left(1+\omega^2 +4 (c^4-c^2)  \right) \; ,\nonumber \\
c^2 &=& \sum_{i=1}^{3} \sin^2 ap_i/2 \;\; , \;\
\omega^2 = \sum_{i=1}^{3} \sin^2 ap_i  \; .
\label{coefficients}
\end{eqnarray} 
The argument of the logarithm may be written as a quadratic form in 
$\xi =\sqrt{c^2x^2+b^2}$, given 
by $\left( \xi^2 + c^2\sqrt{2}\xi + \left(c^2(c^2-1/2)-b^2\right)\right)/(\sqrt{2} c^2 \xi)$. 
Hence  the numerator may be factorized to obtain
\beq
Pa^{4} = \frac{2}{(2\pi)^3} \int_{-\pi}^{\pi} {\rm d}^3 ap
\frac{1}{N_\tau} \sum_{ap_4}\left(
\ln\left(\sqrt{c^2x^2+b^2}-\xi^{+} \right)
+\ln\left(\sqrt{c^2x^2+b^2}-\xi^{-} \right)-
\frac{1}{2}\ln\left(c^2x^2+b^2\right) - \ln\left(\sqrt{2c^4}\, \right)  
\right) \; ,
\label{p3}
\eeq
with $\xi^{\pm} = -c^2/\sqrt{2} \pm \sqrt{b^2-(c^4-c^2)/2}= -c^2/\sqrt{2} \pm 
\sqrt{(1+\omega^2)/8}$. The last term
in Eq.~(\ref{p3}) is easily identified as a vacuum term that gets canceled
once we subtract the zero temperature contribution to the pressure. The third
term has a structure similar to that discussed for staggered fermions. It
corresponds to a dispersion relation that approaches a non-zero value for
small momenta and thus will not contribute to a $1/N_\tau$ expansion  for 
the pressure, {\it i.e.} it only leads to exponentially small contributions.
We thus have to understand the $N_\tau$ dependence of the first two terms.
To do so we first carry out the sum over $p_4$ which can be done in close
analogy to the cases discussed above by first considering the sums
\begin{eqnarray}
\mathcal{S}^{\pm} = \frac{\rm d~~}{{\rm d} \xi^{\pm}} \frac{1}{N_\tau} \sum_{ap_4} 
\ln\left(\sqrt{c^2x^2+b^2}-\xi^{\pm} \right) &=&
- \frac{1}{N_\tau} \sum_{ap_4}  \frac{1}{\sqrt{c^2x^2+b^2}-\xi^{\pm}}  \; .
\label{sum}
\end{eqnarray}
This sum can be evaluated using the same contour integral technique
\cite{Elze} exploited in the case of staggered and Wilson fermions. It,
however, needs a bit more care as through the replacement 
of $x^2=\sin^2(ap_4/2)$
by a complex variable, $x^2 = -(z-1/z)^2/4$, the square root in Eq.~(\ref{sum})
will have cuts. These contribute to the contour integral. Additional
residues appear when doing the sum for $\xi^+$ as 
poles appear in Eq.~(\ref{sum}) for small momenta when $\xi^+ > 0$.  
We discuss the calculation of $\mathcal{S}^{\pm}$ in 
Appendix~\ref{app:overlap} and show there that contributions arising
from the cuts differ from a 
vacuum contribution only by exponentially small terms and thus do not
contribute to the large-$N_\tau$ expansion of the pressure. This expansion
thus receives contributions only from the first term in Eq.~(\ref{sum}) and
only from that part of the momentum integral where  $\xi^+ > 0$.
We thus find,
\begin{eqnarray}
\frac{P}{T^4} \equiv \left[ Pa^4 -(Pa^4)_0\right] N_\tau^4
&=&  \frac{2N_\tau^3}{(2\pi)^3} \int_{d(p)} {\rm d}^3 ap
\sum_{ap_4}
\ln\left(\sqrt{c^2 \sin^2(ap_4/2)+b^2}-\xi^{+} \right) \nonumber \\
& &- \frac{2N_\tau^4}{(2\pi)^4}\int_{d(p)} {\rm d}^4ap 
\ln\left(\sqrt{c^2\sin^2(ap_4/2)+b^2}-\xi^{+} \right) 
  + {\cal O}({\rm e}^{-\alpha N_\tau}) \; ,
\label{ovpressure}
\end{eqnarray}
where $d(p)$ denotes the section of the 3-d Brillouin zone, in which 
$\xi^+ > 0$, {\it i.e.} for which $1+\omega^2 > 4 c^4$ holds. As 
discussed in Appendix~\ref{app:overlap} this result also holds
for non-zero chemical potential, with the replacement
$ap_4\rightarrow ap_4 - i\mu a$.

We are thus left with an expression very similar to what we have discussed
in the previous sections and can express the pressure for overlap fermions
through the poles of the denominator appearing in Eq.~(\ref{sum}),
\begin{equation}
\sinh^2 (E^{+}/2) \equiv
\frac{1}{4}\left(z_i^{+} - \frac{1}{z_i^{+}}\right)^2 = 
\frac{b^2-(\xi^{+})^2}{c^2} = -\frac{1}{2}\left( 1 - \sqrt{1+\omega^2}
\right) \;\;  \Leftrightarrow \;\; \sinh^2 E^+ = \omega^2 \; .
\label{poles}
\end{equation}

We now can integrate Eq.~(\ref{sum}) to arrive at explicit expressions for 
Matsubara sums of logarithms. After
subtracting again the zero temperature contribution to the pressure
one then obtains for the pressure in the overlap formalism an expression 
that agrees with the corresponding result for staggered fermions,
Eq.~(\ref{pressure}) in a limited region of the Brillouin zone that includes
$\vec{p}=0$.Aside from this the pressure for overlap fermions receives 
only contributions that are exponentially small in $N_\tau$. 
This confirms that the large-$N_\tau$ expansion for the pressure in the overlap
formulation agrees with that of staggered fermions.
As indicated at the beginning of this section this could have 
been anticipated, for we know that the two actions have the same dispersion 
relation at small momenta. 

\begin{figure}
\epsfig{file=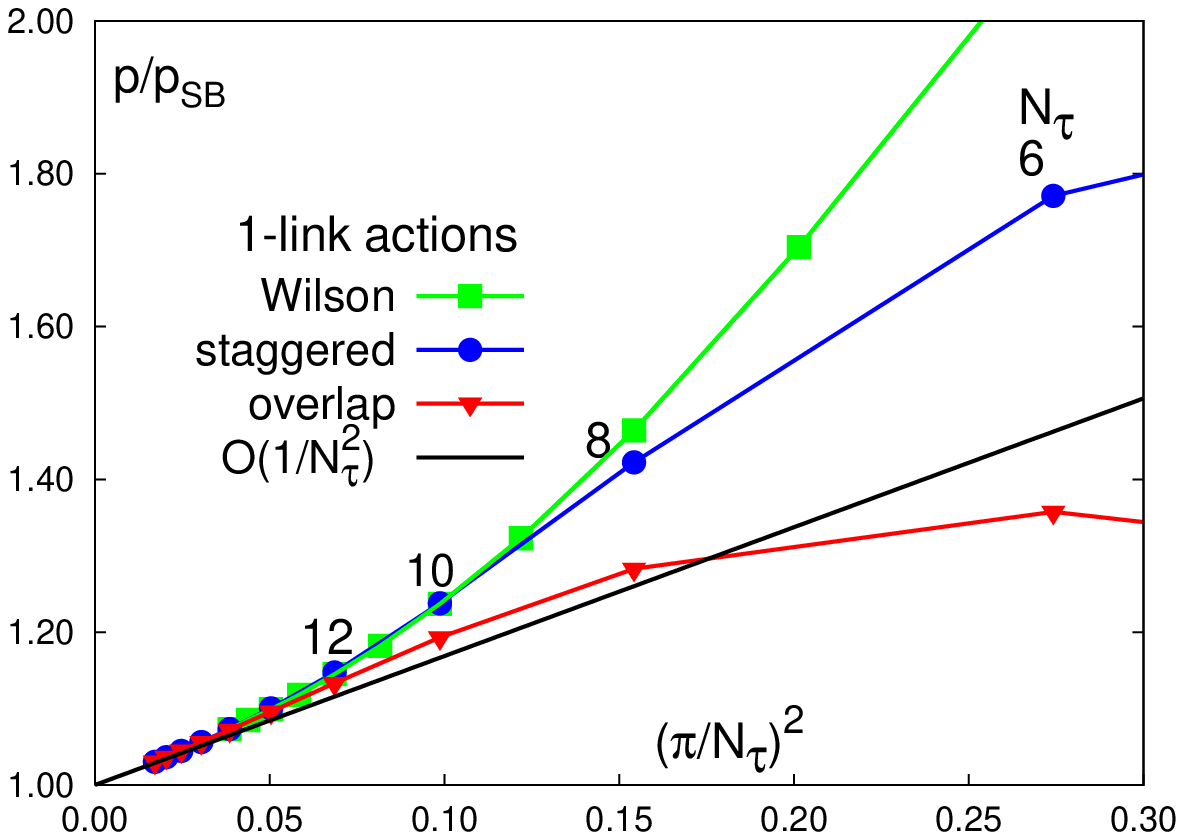,width=7.5cm}
\epsfig{file=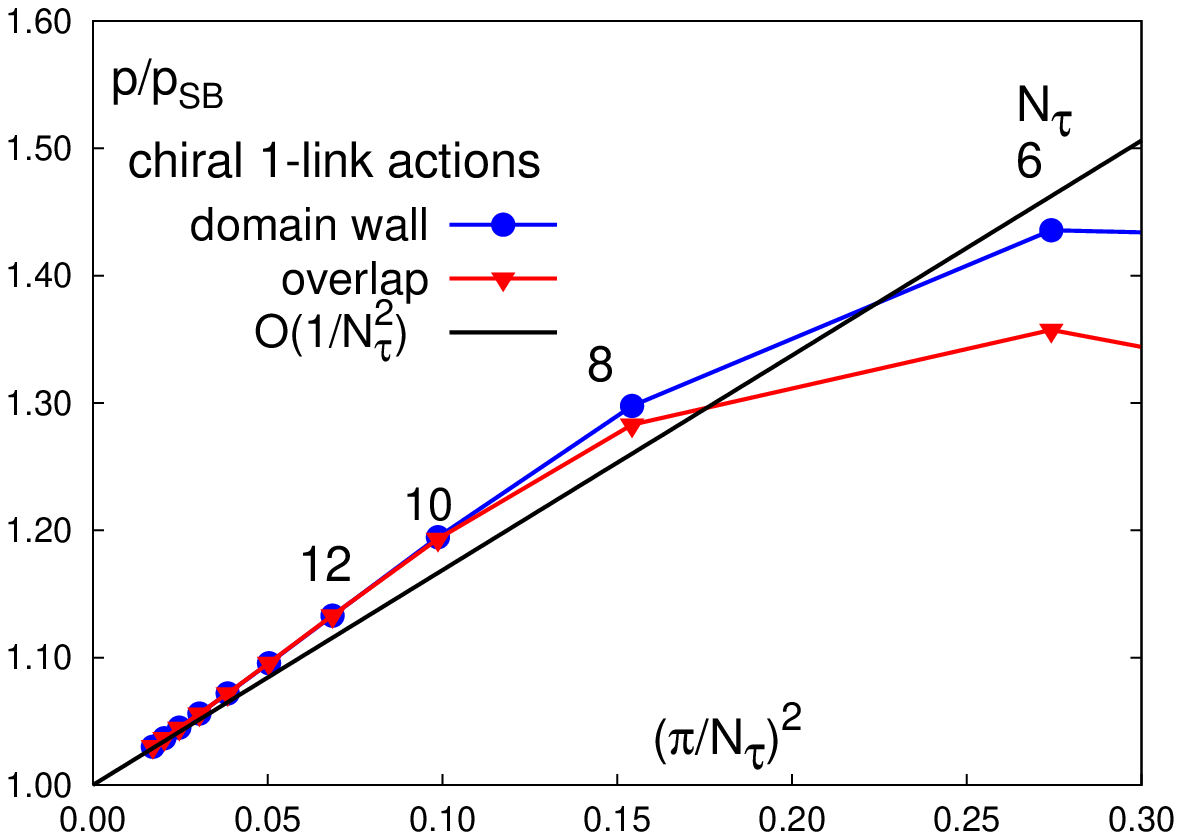,width=7.5cm}
\caption{Pressure for overlap fermions based on the application
of the Ginsparg-Wilson operator on a standard Wilson operator (left) and
for the domain wall operator using the same kernel with domain wall height
$M_5=1$(right).
The large-$N_\tau$ expansion for both discretization schemes 
coincides with that of the standard staggered action. In both
figures we compare with the leading order $1/N_\tau^2$ result.
 }
\label{fig:chiral}
\end{figure}

For fixed values of $N_\tau$ the exact expression for the pressure, 
Eq.~(\ref{opres}), of course, differs from that of naive staggered fermions
and cut-off effects thus will be different. Using Eq.~(\ref{opres}) we have 
evaluated the pressure for small values of $N_\tau$ explicitly.
The results are compared to corresponding calculations performed for
standard staggered and Wilson fermions as well as to the common leading
order large-$N_\tau$ expansion result in Fig.~\ref{fig:chiral}(left).
We note that the cut-off dependence introduced by the overlap operator 
for small values of $N_\tau$ is smaller than in the staggered fermion
discretization scheme. In fact, it is remarkable that already for
$N_\tau =8$ the exact result for overlap fermions is close to the 
leading order result of the large-$N_\tau$ expansion. For
$N_\tau \gsim 12$ all 1-link discretization schemes yield very similar
results and are in good agreement with the leading order large-$N_\tau$ 
expansion.

We finally comment on the ideal gas limit for domain wall fermions.
This has previously been analyzed numerically on finite space-time lattices
\cite{Fleming}. In the analysis of the cutoff dependence of bulk 
thermodynamics with domain wall fermions an additional complication arises.
One has to introduce additional degrees of freedom, Pauli-Villar regulators,
which eliminate divergences that arise from the infinite number of heavy
fermion fields present in the limit of infinite 5th dimension
\cite{Neuberger2}.
The calculation of the free fermion partition function for
domain wall fermions, {\it i.e.} the free fermion determinant, thus is
somewhat more involved \cite{Edwards}. Nonetheless, after diagonalization
in momentum space  the regularized domain wall fermion determinant 
takes on a simple form in the limit of infinite extent of the
5th dimension. For massless fermions one obtains,
\begin{equation}
{\rm det} D_{DW} = \prod_{\vec{p},p_4} {\cal N}(\vec{p},p_4) \; , 
\label{detDW}
\end{equation}
with
\begin{equation}
{\cal N}(\vec{p},p_4) = 2 \left( 1 + 
\frac{S^2+ W^2 -1}{\sqrt{(1+S^2 + W^2)^2 - 4 W^2}} \right) \; .
\label{NDW}
\end{equation}
Here $S^2$ has been defined in Eq.~(\ref{AS}) and 
$W = 1-M_5 - 2 \sum_{\nu=1}^{4} \sin^2(ap_\nu /2)$ for domain wall height
$M_5$.

As mentioned before also domain wall fermions have the same dispersion
relation as staggered fermions, up to a constraint, which cuts off the
spectrum for large momenta. The large-$N_\tau$ expansion of bulk
thermodynamics in the ideal gas limit thus again will be identical to
that of naive staggered fermions. Moreover, for vanishing lattice spacing 
in the 5th dimension and infinite extent in this direction 
the domain wall Dirac operator is identical to the overlap 
operator. It thus may not be surprising that the cut-off effects in
the domain wall and in the overlap formalism are quite similar in the free
field limit. Differences arise because the sign function is 
introduced in the domain wall formalism through a discrete set of additional 
fermion fields \cite{Neuberger2}. We limit our discussion of cutoff
effects in thermodynamics with domain wall fermions here to a 
comparison of the exact evaluation of the pressure in the domain wall
and overlap discretization schemes. With Eqs.~(\ref{detDW}) and (\ref{NDW})  
the pressure of free massless domain wall fermions is obtained as  
\begin{eqnarray}
\frac{P}{T^4} \equiv \left[ Pa^4 -(Pa^4)_0\right] N_\tau^4
&=&  \frac{2 N_\tau^3}{(2\pi)^3} \int_{-\pi}^{\pi} {\rm d}^3 ap
\sum_{ap_4}
\ln {\cal N}(\vec{p},p_4) 
- \frac{2 N_\tau^4}{(2\pi)^4}\int_{-\pi}^{\pi} {\rm d}^4ap 
\ln  {\cal N}(\vec{p},p_4) 
 \; .
\label{dwfpressure}
\end{eqnarray}
In Fig.~\ref{fig:chiral}(right)
we show results from a direct evaluation of this expression for 
some values of $N_\tau$ and for domain wall height $M_5=1$, which corresponds
to the case discussed above for overlap fermions. 
As can be seen cutoff effects differ for
small values of $N_\tau$ but rapidly become similar for $N_\tau \simeq 8$.

\section{Conclusions}
\label{se:conclusion}

We have analyzed here cut-off effects that arise in lattice 
calculations of bulk thermodynamic observables at high temperature
and vanishing as well as non-vanishing chemical potential. We have
shown that in the high temperature, ideal gas limit of QCD cut-off effects
in fermionic observables can be traced back
to the cut-off dependence of the dispersion relation for the quark
propagator. The structure of the cut-off dependence is preserved at
non-zero chemical potential $\mu$, {\it i.e.} actions which are improved
to a certain order in $aT\equiv 1/N_\tau$ at $\mu=0$ are so also for
$\mu > 0$. We could show that the $\mu$-dependence
of cut-off effects is universal, {\it i.e.} $\mu$-dependent correction factors
at ${\cal O}(N_\tau^n)$ are independent of the discretization and improvement
schemes. They are given in terms of a Bernoulli polynomial of degree $4+n$;
cut-off corrections at order $1/N_\tau^n$  depend on an even 
polynomial in $\mu/\pi T$ of degree $4+n$ with positive coefficients only.

We have shown that the considerations on cut-off effects in thermodynamics 
with staggered and Wilson fermions carry over to chiral fermion formulations.
So far the latter have been used mainly with standard Wilson fermion kernels.
The discussion presented in this paper suggest that studies of bulk
thermodynamics performed with chiral fermions could profit a lot from
the application of improvements schemes similar to those used in 
studies of bulk thermodynamics with staggered fermions. This certainly
can be achieved using the concepts that have been developed to improve
standard staggered and Wilson fermion action \cite{Christ}. 

\section*{Acknowledgments}
\label{ackn}
\noindent
We thank Urs Heller for helpful discussions.
SS has been supported by the DFG grant GRK/881 and by the EU under 
the contract no. RII3-CT-2004-506078. 
FK and EL acknowledge partial support through 
a grant of the BMBF under contract no. 06BI106.
The work of FK and PH has been supported by a contract DE-AC02-98CH10886 with
the U.S. Department of Energy. 

\begin{appendix}
\renewcommand{\thefigure}{D.\arabic{figure}}
\renewcommand{\thetable}{D.\arabic{figure}}
\setcounter{figure}{0}
\setcounter{table}{0}
\section{Large \boldmath$N_\tau$ expansion of the pressure}
\label{app:pressure}

We give here some further details on the derivation of the series expansion
of the pressure, Eq~(\ref{pressureNtk}), and give explicit formulas for the
expansion coefficients $A_n$ as well as the polynomials $P_n$ introduced in
Eq.(\ref{pressureNtk}).

In order to derive Eq.~(\ref{pressureTaylor}) from Eq.~(\ref{pressureNt}) as
the result of a Taylor expansion $y\Delta$ it is helpful to note that
\begin{equation}
\frac{\partial \ln [1-B(y,z) (1-{\rm e}^{-y\Delta})]}{\partial y\Delta}
= \frac{-1}{z \exp(y) \exp(y\Delta) +1} \;  .
\label{derivative}
\end{equation}
It then is obvious that taking further partial derivatives of this logarithm 
with respect to $y\Delta$ at $y\Delta=0$ is equivalent to taking 
derivatives with respect to $y$ at $y\Delta=0$. This yields 
Eq.~(\ref{pressureTaylor}).

The series representation of the pressure, Eq~(\ref{pressureNtk}), is then
obtained from Eq.~(\ref{pressureTaylor}) after noting that the expansion
coefficient at order $1/N_\tau^{2k}$ receives contributions from all terms
in Eq.~(\ref{pressureTaylor}) with $n\le k$  as $\Delta$ is a polynomial
in $y/N_\tau$ starting at ${\cal O}(1/N_\tau^2)$.  Aside from a complicated 
factor that only depends on the angular variables $(\phi,\; \theta)$,
these contributions are proportional to $y^{2k}$. Irrespective of the structure
of the dispersion relation of a particular action under consideration the 
contribution of the radial integral in Eq.~(\ref{pressureTaylor}) to an
expansion coefficient at order $1/N_\tau^{2k}$ thus has the unique form 
\begin{eqnarray}
\int_0^\infty {\rm d} y y^{2k+2+n}
\left( \frac{\partial^{n-1}B(y,z)}{\partial y^{n-1}} +
\frac{\partial^{n-1}B(y,z^{-1})}{\partial y^{n-1}}\right) &=&
(-1)^{n-1}\frac{(2k+2+n)!}{(2k+3)!} \int_0^\infty {\rm d} y y^{2k+3} 
\left( B(y,z) + B(y,z^{-1}) \right) \nonumber \\
&=& (-1)^{n}\frac{(2k+2+n)!}{(2k+4)!} (2i \pi)^{2k+4}
B_{2k+4}\left(\frac{\mu}{2\pi T i}+\frac{1}{2} \right) 
\label{ya}
\end{eqnarray}
The angular contribution is given by
\begin{eqnarray}
A_{2k}&=&\frac{b_{4+2k}}{4\pi^3}
\sum_{n=1}^{k}\frac{1}{n!} 
\int_0^{\pi}{\rm d} \theta \sin \theta \int_0^{2 \pi}{\rm d}\phi 
\sum_{(k_1,..k_n)}\prod_{j=1}^{n}  
a_{k_j}(\phi,\theta) \; ,
\label{angular}
\end{eqnarray}
where the inner sum extends over all n-tuples of positive even
integers with $\sum_{j=1}^n k_j = 2k$. Here we have introduced a 
multiplicative normalization factor, the Bernoulli number $b_{4+2k}$,
that arises from the integration of the radial variable $y$ at vanishing
chemical potential and normalizes the polynomials $P_{2k}(x)$ to unity
for $x=0$.

For a few small values of the expansion parameter $k$
we give here the explicit form of the polynamials $P_{2k}(x)$ defined in 
Eq.~(\ref{Bernoulli}):
\begin{eqnarray}
P_0(x) &=& 1+ \frac{30}{7} x^2 +\frac{15}{7} x^4 \nonumber \\
P_2(x) &=& 1+ \frac{147}{31} x^2 +\frac{105}{31} x^4 + \frac{21}{31} x^6 \nonumber \\
P_4(x) &=& 1+ \frac{620}{127} x^2 +\frac{490}{127} x^4 + \frac{140}{127} x^6
+\frac{15}{127} x^8 \nonumber \\
P_6(x) &=& 1+ \frac{12573}{2555} x^2 +\frac{2046}{511} x^4 + \frac{462}{365} x^6
+\frac{99}{511} x^8 +\frac{33}{255} x^{10} 
\end{eqnarray}

\section{The propagator for hypercube Wilson fermions}
\label{app:hypercube}

using the shorthand notation, $c_i = \cos ap_i$ and $s_i=\sin ap_i$
we write the the denominator of the fermion propagator for 
Wilson type fermions, Eq.~(\ref{eq:Dwil}), as
\beq
D(p_4, \vec p) =
(\frac{1}{4} {\rm Tr} {\cal K}_1^2 + \kappa_1^2 + \delta^2)
+ 2 \cos k_4 (\frac{1}{4} {\rm Tr} {\cal K}_1 {\cal K}_2 + \kappa_1 \kappa_2)
+ \cos^2 k_4 (\frac{1}{4} {\rm Tr} {\cal K}_2^2 + \kappa_2^2 - \delta^2)
\eeq
where
\bs
\kappa_1(\vec p) & = &
\lambda_0 + 2 \lambda_1 (c_1 + c_2 + c_3) + 
4 \lambda_2 (c_1 c_2 + c_2 c_3 + c_3 c_1) + 8 \lambda_3 c_1 c_2 c_3 \\
\kappa_2(\vec p) & = &
2 \lambda_1 + 4 \lambda_2(c_1 + c_2 + c_3) + 
8 \lambda_3 (c_1 c_2 + c_2 c_3 + c_3 c_1) + 16 \lambda_4 c_1 c_2 c_3 \\
\delta(\vec p) & = &
2 \rho_1 + 4 \rho_2 (c_1 + c_2 + c_3) + 
8 \rho_3 (c_1 c_2 + c_2 c_3 + c_3 c_1) + 16 \rho_4 c_1 c_2 c_3 
\es
and the ${\cal K}_i$ matrices are given by
\beq
{\cal K}_1 = \sum_{i=1}^3 \gamma_i \alpha_i \;  ~~~~
{\cal K}_2 = \sum_{i=1}^3 \gamma_i \beta_i  \; .
\eeq
The coefficients appearing in this matrix representation are,
\bs
\alpha_1(\vec p) & = & 2 s_1 ( \rho_1 + 2 \rho_2 (c_2 + c_3) + 4 \rho_3 c_2 c_3 ) \\
\alpha_2(\vec p) & = & 2 s_2 ( \rho_1 + 2 \rho_2 (c_3 + c_1) + 4 \rho_3 c_3 c_1 ) \\
\alpha_3(\vec p) & = & 2 s_3 ( \rho_1 + 2 \rho_2 (c_1 + c_2) + 4 \rho_3 c_1 c_2 ) \\
 & & \nonumber \\
\beta_1(\vec p) & = & 4 s_1 ( \rho_2 + 2 \rho_3 (c_2 + c_3) + 4 \rho_4 c_2 c_3 ) \\
\beta_2(\vec p) & = & 4 s_2 ( \rho_2 + 2 \rho_3 (c_3 + c_1) + 4 \rho_4 c_3 c_1 ) \\
\beta_3(\vec p) & = & 4 s_3 ( \rho_2 + 2 \rho_3 (c_1 + c_2) + 4 \rho_4 c_1 c_2 ) 
\es
Finally, the expressions $P, Q$ and $R$ used in Eq.(\ref{eq:Dwil}) and
(\ref{eq:wil_e}) are obtained as
\beq
P(\vec p) = \frac{1}{4} {\rm Tr} {\cal K}_1 {\cal K}_2 + \kappa_1 \kappa_2 ~~~~
Q(\vec p) = \frac{1}{4} {\rm Tr} {\cal K}_2^2 + \kappa_2^2 - \delta^2 ~~~~
R(\vec p) = \frac{1}{4} {\rm Tr} {\cal K}_1^2 + \kappa_1^2 + \delta^2
\eeq

\section{The expansion coefficient \boldmath $h_6$}
\label{app:fp6}

We here give the explicit result for the sixth order expansion coefficient 
$h_6$ appearing in Eq.~(\ref{Wilson_P}), which has been obtained using 
Mathematica. It reads
\begin{eqnarray}
h_6 &=&\frac{-73}{8316}
(-2288 + 734400 \lambda_3^2  + 470016 \lambda_4^2 + (1175040 - 2585088 r + 1410048 \rho^2) \lambda_3 \lambda_4
\nonumber\\
&&\hspace{0.9cm}+ 82584 \rho_2 - 918432 \rho_2^2  + 288000 \rho_3 - 6846336 \rho_2 \rho_3 - 12690432 \rho_3^2  + 247392 \rho_4
\nonumber \\
&&\hspace{0.9cm}- 6345216 \rho_2 \rho_4 - 23376384 \rho_3 \rho_4 - 10685952 \rho_4^2  +  
29376 \lambda_2^2  (2 - 3 r)^2  + 1836 \lambda_1^2  (1 - 2 r)^2  
\nonumber \\
&&\hspace{0.9cm}- 1762560 \lambda_3^2  r - (940032 - 470016 r) \lambda_4^2 r + 1057536 \lambda_3^2  r^2  + 5967 r^3 
\nonumber \\
&&\hspace{0.9cm}+ 58752 \lambda_2 (-2 + 3 r) (4 \lambda_4 (-1 + r) + \lambda_3 (-5 + 6 r)) 
\nonumber \\
&&\hspace{0.9cm}+ 14688 \lambda_1 (-1 + 2 r) (4 \lambda_4 (-1 + r) + \lambda_2 (-2 + 3 r) + \lambda_3 (-5 + 6 r)))
\label{h6}
\end{eqnarray}
where we have introduced the short hand notation,
\beq
r=8 \rho_2+ 32 \rho_3 + 32 \rho_4 \; .
\eeq

\section{Matsubara sums for overlap fermions}
\label{app:overlap}

We give here some further details on the evaluation of Matsubara
sums appearing in calculations with overlap fermions. 
Throughout this appendix we make the dependence on the quark chemical 
potential explicit. We want to evaluate
the sum
\begin{equation}
\mathcal{S(\alpha)} = \frac{1}{N_\tau} \sum_{p_4}
\ln\left(\sqrt{c^2x^2+b^2}-\alpha\right)
\end{equation}
where $x=\sin((ap_4-i\mu a)/2)$, and $b$, $c$ and $\alpha$ are functions of the
three-momenta $(p_1,p_2,p_3)$ and $ap_4=2\pi (n+1/2)/N_\tau$ with
$n=0,\pm 1,...,\pm (N_\tau-1)/2, N_\tau/2$.

\begin{figure}[t]
\begin{center}
\epsfig{file=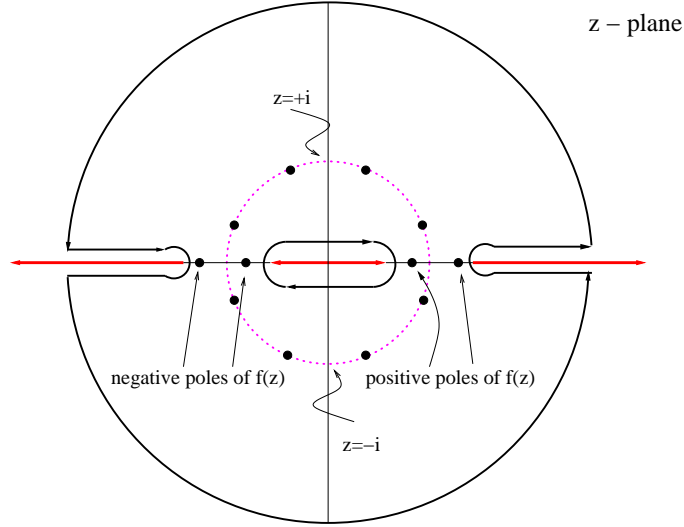,width=9.0cm}
\caption{The locations of the poles and branch cuts of $H(z)$ and the choice 
of contour for $N_\tau=4$. The four poles of $f(z)$ only exist for $\alpha >0$.
Also shown is the unit circle (dotted line) and the location of poles arising
from the exponential integration kernel introduced in Eq.~(\ref{Hz}).}
\end{center}
\label{fig:cont}
\end{figure}

To evaluate it, we first differentiate with respect to $\alpha$,
\begin{equation}
\frac{\mathrm{d}\mathcal{S}}{\mathrm{d}\alpha}=-\frac{1}{N_\tau}
\sum_{p_4}\frac{1}{\sqrt{c^2x^2+b^2}-\alpha} \; . 
\label{sumsqrt}
\end{equation}
This sum can be thought of as being part of the result of a contour integral 
over the function,
\begin{eqnarray}
H(z)= \frac{1}{\sqrt{-c^2\left(z-\frac{1}{z}\right)^2/4 +b^2}
-\alpha} \;
\frac{\mathrm{e}^{\mu aN_\tau}}{z \left(z^{2 N_\tau}+\mathrm{e}^{\mu aN_\tau}
\right)}
= f(z) 
\frac{\mathrm{e}^{\mu aN_\tau}}{z (z^{2 N_\tau}+\mathrm{e}^{\mu aN_\tau})}\; .
\label{Hz}
\end{eqnarray}
where the second equality defines the function $f(z)$.
The function $H(z)$ has poles and cuts; the sum given in Eq.~(\ref{sumsqrt})
is related to the poles on the unit circle which arise from the last factor 
in the denominator of Eq.~(\ref{Hz}). We want to evaluate the contour integral
along the path shown in Fig.~\ref{fig:cont}.1.
As long as $\alpha \equiv \alpha(p)$ is positive there are four additional
poles arising from $f(z)$ which are located at $\pm z_0$ and $\pm 1/z_0$ with,
\begin{equation}
z_0 = e^{\tilde{x}/2} \;\;\; {\rm and}\;\;\; \cosh\tilde{x}=1+\frac{2}{c^2}
\left(b^2-\alpha^2\right) \; . 
\end{equation}
The residues at these points are listed in Table~\ref{tab:residues}.1.
\begin{table}[t]
\begin{center}
\begin{tabular}{|r|c|}
\hline
Pole: $z$~ & Residue of $H(z)$ \\
\hline
~&~\\[-6pt]
~$z_0$~ & $\displaystyle{\frac{2 \alpha}{z_0^2-(1/z_0)^2} 
\cdot\frac{1}{(z_0^2\mathrm{e}^{-\mu a})^{N_\tau}+1}}$ \\[12pt]
$-z_0$~ & $\displaystyle{\frac{2 \alpha}{z_0^2-(1/z_0)^2}
\cdot\frac{1}{(z_0^2\mathrm{e}^{-\mu a})^{N_\tau}+1}}$ \\[12pt]
~$1/z_0$~ & $\displaystyle{\frac{-2 \alpha}{z_0^2-(1/z_0)^2} 
\cdot\frac{1}{(z_0^{-2}\mathrm{e}^{-\mu a})^{N_\tau}+1}}$ \\[12pt]
$-1/z_0$~ & $\displaystyle{\frac{-2 \alpha}{z_0^2-(1/z_0)^2}
\cdot\frac{1}{(z_0^{-2}\mathrm{e}^{-\mu a})^{N_\tau}+1}}$ \\[12pt]
\hline
\end{tabular}
\caption{The four residues of $H(z)$ at the poles of $f(z)$ which exist for
$\alpha > 0$.}
\end{center}
\label{tab:residues}
\end{table}

Moreover, there are branch cuts arising from the square root, which we take 
to be located in the intervals $C^-=(-\infty, w_0)$, $C^0=(-1/w_0,1/w_0)$, 
$C^+= (w_0,\infty)$, with $\sinh(w_0)= b/c$. We note that the contributions
from cuts with ${\rm Re}z>0$ are identical to those with ${\rm Re}z<0$.
Moreover, one can map the contribution from the positive part of the central
cut, $(0,1/w_0)$, onto the interval $C^+$. With this we arrive at the 
result,
\begin{eqnarray}
\frac{\mathrm{d}\mathcal{S}}{\mathrm{d}\alpha} =
&&-\Theta(\alpha) 
\frac{4 \alpha}{c^2\sinh \tilde{x}}
\left(
\frac{1}{\mathrm{e}^{-N_\tau(\tilde{x}+\mu a)}+1}
-\frac{1}{\mathrm{e}^{N_\tau(\tilde{x}-\mu a)}+1}
\right) \nonumber \\
&&-\frac{2}{\pi} \mathrm{Im} \int_{w_0}^\infty 
\frac{f(z)}{z}
\left(
\frac{1}{(z^2 \mathrm{e}^{-\mu a})^{N_\tau} + 1}  
-\frac{1}{(z^{-2} \mathrm{e}^{-\mu a})^{N_\tau} + 1} 
\right)\; . 
\end{eqnarray}
Integrating this again we obtain
\begin{eqnarray}
\mathcal{S(\alpha)} = &&\Theta(\alpha) \frac{1}{N_\tau}
\left(
\ln (1 + \mathrm{e}^{-N_\tau (\tilde{x}-\mu a)})
+\ln (1 + \mathrm{e}^{-N_\tau (\tilde{x}+\mu a)}) +N_\tau \tilde{x}
\right) \nonumber \\
&&+
\frac{2}{\pi} \mathrm{Im} \int_{w_0}^\infty {\rm d}z
\frac{\ln\left( \sqrt{-c^2\left(z-\frac{1}{z}\right)^2/4 +b^2} -
\alpha \right)}{z}
\left(
\frac{1}{(z^2 \mathrm{e}^{-\mu a})^{N_\tau} + 1}  
-\frac{1}{(z^{-2} \mathrm{e}^{-\mu a})^{N_\tau} + 1} 
\right)  
+ {\rm const.}
\label{total}
\end{eqnarray}
It easily is seen that the absolute value of the integral appearing here is 
bounded
by ${\rm const} \cdot \exp(-N_\tau w_0)$. When we insert $S(\alpha)$
in Eq.~(\ref{ovpressure}) we thus find that only the first term in 
Eq.~(\ref{total})
contributes to the large-$N_\tau$ expansion of the pressure. 
\end{appendix}


\begin{thebibliography}{99}
\bibitem{Symanzik}
K.~Symanzik, Nucl.\ Phys.\  B {\bf 226}, 187 (1983) and
Nucl.\ Phys.\  B {\bf 226}, 205 (1983).
\bibitem{Naik} S. Naik, Nucl. Phys. B {\bf 316}, 238 (1989).
\bibitem{p4action} F. Karsch, E. Laermann and A. Peikert, Nucl. Phys.
         B {\bf 605}, 579 (2001).
\bibitem{clover}
B. Sheikholeslami and R. Wohlert, Nucl. Phys. B {\bf 259}, 572 (1985).
\bibitem{Hasenfratz}
P.~Hasenfratz and F.~Niedermayer,
Nucl.\ Phys.\  B {\bf 414}, 785 (1994);\\
P.~Hasenfratz, S.~Hauswirth, K.~Holland, T.~Jorg, F.~Niedermayer and U.~Wenger,
Int.\ J.\ Mod.\ Phys.\  C {\bf 12}, 691 (2001).
\bibitem{Bietenholz} W. Bietenholz and U.-J. Wiese,
         Nucl. Phys. B {\bf 464}, 319 (1996).
\bibitem{Boyd}
G.~Boyd, J.~Engels, F.~Karsch, E.~Laermann, C.~Legeland, M.~Lutgemeier and 
B.~Petersson,
Nucl.\ Phys.\  B {\bf 469}, 419 (1996).
\bibitem{Beinlich}
B.~Beinlich, F.~Karsch, E.~Laermann and A.~Peikert,
Eur.\ Phys.\ J.\  C {\bf 6}, 133 (1999).
\bibitem{Allton} 
C.~R.~Allton, S.~Ejiri, S.~J.~Hands, O.~Kaczmarek, F.~Karsch, E.~Laermann 
and C.~Schmidt,
Phys.\ Rev.\  D {\bf 68}, 014507 (2003).
\bibitem{milc} 
C.~Bernard {\it et al.}  [MILC Collaboration],
Phys.\ Rev.\  D {\bf 71}, 034504 (2005) and
Phys.\ Rev.\  D {\bf 77}, 014503 (2008).
\bibitem{perfectmu}
W. Bietenholz and U.-J. Wiese, Phys. Lett. B {\bf 426}, 114 (1998).
\bibitem{Gattringer}
C.~Gattringer and L.~Liptak,
Phys.\ Rev.\  D {\bf 76}, 054502 (2007).
\bibitem{Hasenfratzmu}
P.~Hasenfratz and F.~Karsch,
Phys.\ Lett.\  B {\bf 125}, 308 (1983).
\bibitem{Elze} H.-Th. Elze, K. Kajantie and J. Kapusta,
Nucl. Phys. B {\bf 304}, 832 (1988).
\bibitem{perfectnew}
W. Bietenholz, R. Brower, S. Chandrasekharan and  U.-J. Wiese,
Nucl. Phys. Proc. Suppl. {\bf 53}, 921 (1997).
\bibitem{Shcheredin}
S.~Shcheredin and E.~Laermann,
PoS {\bf LAT2006}, 146 (2006).
\bibitem{Adams} D.H. Adams, Nucl. Phys. (Proc. Suppl.) {\bf 129\&130}, 513 
(2004).
\bibitem{Neuberger}
R. Narayanan and H. Neuberger, Phys. Lett. B {\bf 302}, 62 (1993) and
Nucl. Phys. B {\bf 443}, 305 (1995).
\bibitem{Shamir}
Y. Shamir, Nucl. Phys. B {\bf 406}, 90 (1993).
\bibitem{Kaplan}
D. Kaplan, Phys. Lett. B {\bf 417}, 141 (1998).
\bibitem{Capitani}
for a recent review see: S. Capitani, Phys. Rept. {\bf 382}, 113 (2003).
\bibitem{Aarts}
G.~Aarts and J.~Foley  [UKQCD Collaboration],
JHEP {\bf 0702}, 062 (2007).
\bibitem{Fleming}
G.~T.~Fleming,
  Nucl.\ Phys.\ Proc.\ Suppl.\  {\bf 94}, 393 (2001).
\bibitem{Neuberger2}
H.~Neuberger,
  Phys.\ Rev.\  D {\bf 57}, 5417 (1998)
\bibitem{Edwards}
R.~G.~Edwards and U.~M.~Heller,
Phys.\ Rev.\  D {\bf 63}, 094505 (2001).
\bibitem{Christ}
N. Christ, private communication
\end{thebibliography}
\end{document}